# Real-time Detection of Clustered Events in Video-imaging data with Applications to Additive Manufacturing

Hao Yan[1*], Marco Grasso[2], Kamran Paynabar[3], and Bianca Maria Colosimo[2]

[1] *School of Computing, Informatics, and Decision Systems Engineering, Arizona State University, Tempe, AZ, USA,*

[2] *Dipartimento di Meccanica, Politecnico di Milano, Italy*

[3] *H. Milton Stewart School of Industrial and Systems Engineering, Georgia Institute of Technology, Atlanta, GA, USA*

* corresponding author: haoyan@asu.edu

*Abstract.* The use of video-imaging data for in-line process monitoring applications has become more and more popular in the industry. In this framework, spatio-temporal statistical process monitoring methods are needed to capture the relevant information content and signal possible out-of-control states. Video-imaging data are characterized by a spatio-temporal variability structure that depends on the underlying phenomenon, and typical out-of-control patterns are related to the events that are localized both in time and space. In this paper, we propose an integrated spatio-temporal decomposition and regression approach for anomaly detection in video-imaging data. Out-of-control events are typically sparse spatially clustered and temporally consistent. Therefore, the goal is to not only detect the anomaly as quickly as possible ("when") but also locate it ("where"). The proposed approach works by decomposing the original spatio-temporal data into random natural events, sparse spatially clustered and temporally consistent anomalous events, and random noise. Recursive estimation procedures for spatio-temporal regression are presented to enable the real-time implementation of the proposed methodology. Finally, a likelihood ratio test procedure is proposed to detect when and where the hotspot happens. The proposed approach was applied to the analysis of video-imaging data to detect and locate local over-heating phenomena ("hotspots") during the layer-wise process in a metal additive manufacturing process.

## 1. Introduction

Nowadays, the use of spatio-temporal data streams such as images and videos in change detection and process monitoring is becoming more popular in advanced manufacturing systems (Megahed & Jones-Farmer, 2015; Megahed et al., 2011) and other complex systems. On the one hand, the increasing availability of compact, low-cost, and robust machine vision systems that can be easily integrated into production plants has enabled real-time image acquisition. On the other hand, continuously improving computational capabilities has made an in-line analysis of image streams more feasible. An effective anomaly detection method for such data streams should address the following challenges: 1) *High dimensionality*: high-resolution images are comprised of thousands and millions of pixels; 2) *High velocity*: a standard video camera collects 24 frames per second, whereas a high-speed camera may acquire thousands of frames per second, which requires real-time analysis of image frames. 3) *No anomaly labels:* In most of the industrial applications, there are few anomaly samples presented and normally no labels are provided for whether the sample is anomalous or not. 4) *Complex spatio-temporal correlation structure*: neighbor pixels are spatially correlated, and consecutive image frames are temporally correlated.

One specific challenge that this paper addresses is to detect *spatio-temporal correlated anomalies* from random foreground events. Such examples are very common in different applications. In disease monitoring, the foreground events are individual disease infections, where the anomaly events are the systematic disease infection events that are spatially clustered and temporally consistent (Unkel et al., 2012). In crime monitoring (Rogerson & Sun, 2001), the foreground events are typically random crime events happening in different places, where the anomaly event is the local and clustered hotspots with more crime events happening. In the defect detection in complex systems, the foreground events are the random failures in the system and the anomaly events are the persistent and clustered failure (Yan et al., 2018).

This paper presents a new scalable spatio-temporal decomposition methodology to detect the structured anomalies in real time. In this study, we made the following common assumptions about the spatial and temporal structure of foreground natural events and anomaly events: 1) the foreground natural events are sparse and random in the spatio-temporal domain. 2) the anomaly event is sparse spatially clustered and temporally consistent. In this paper, we will focus on a case study in the video monitoring example for the additive manufacturing, where the anomaly represents an involving connected pixels (i.e., spatially clustered) that exhibit high pixel intensities for a certain period of time (i.e., temporally consistent). It is worth noting that as long as these assumptions hold, the proposed methodology is applicable to any spatio-temporal process monitoring/change detection. Furthermore, we propose an efficient and recursive estimation procedure to detect and locate the anomaly event in real-time, i.e., as soon as a new data point (e.g., video frame) is recorded. In order to automatically alarm in the occurrence of an anomaly event, we propose to combine the penalized spatio-temporal regression framework with a likelihood ratio test (LRT) for change detection, process monitoring, and anomaly localization (Gertler, 2017; Mood, 1950).

### 1.1 A Motivating Case Study

In the recent years, particular interest has been devoted to the use of machine vision in metal additive manufacturing (AM) applications (Everton et al., 2016; Grasso and Colosimo, 2017; Mani et al., 2015; Spears and Gold, 2016; Tapia and Elwany, 2014). Indeed, the layer-wise production paradigm involved in AM allows one to acquire images and videos during the production of each layer. This yields the capability of measuring several quantities that are proxies of the part quality and the process stability, while the part is being produced, and provides with several benefits including defect detection, waste reduction and cost savings in post-process inspection. In-situ and in-line monitoring of manufacturing processes based on video-imaging data require the capability of making sense of big data streams in an efficient and sound way.

The motivating case study is dealing with in-situ defect detection in laser powder bed fusion (LPBF). LPBF is a metal AM approach where a laser beam is used to selectively melt a powder bed (Stucker et al., 2010). Despite the great industrial potentials of LPBF technology, various kinds of defects may originate during the process (Everton et al., 2016; Grasso & Colosimo, 2017). Nowadays, most industrial LPBF systems are equipped with sensors suitable to measure several quantities during the process (Grasso and Colosimo, 2017), but what is still lacking in industry is the availability of analytical tools able to quickly make sense of gathered data during the process and automatically signal the onset of defects and process instabilities. In this framework, in-situ video imaging allows one to monitor the stability of the process while the part is being produced on a layer-by-layer basis and to detect the onset of process defects. However, although such defects are visually detectable from image streams, automatic and real-time analysis of images is imperative for a scalable and effective process monitoring. The main goal of this paper is to develop an online real-time analytical method for detecting and locating over-heating phenomena in LPBF known as “hot-spots” via in-situ video-imaging. A hot-spot is a region of the powder bed where local heat accumulation occurs because of the excessive energy input and a diminished heat flux towards the surrounding material (Colosimo & Grasso, 2018; Grasso et al., 2017). The quick detection and localization of hot-spots is a key issue in the reduction of scrap fractions in LPBF, as hot-spots may lead to local geometrical distortions and microstructural inhomogeneity in the manufactured part.

Figure 1 shows an example of video-frame acquired during the LPBF of a metal part. The dark area corresponds to the background, where no action occurs. The foreground region, however, includes: i) the natural process including the laser-heated zone (LHZ), i.e., the high-intensity region including and surrounding the zone where powder melting occurs, and the spatters generated by the laser-material interaction, and ii) a hot-spot, i.e., the anomaly to be detected. The LHZ displaces along the pre-defined scanning path of the laser, and its size mainly depends on the energy input. Spatters consist of either hot particles of the powder bed blown away by the metallic vapor or molten material ejected by the melt pool (Khairallah et al., 2016; Y. Liu et al., 2015). Therefore, we can claim that hot spot is sparse and clustered in the spatial domain, and when a hot spot happens, it stays at the same location for some time. However, the spatters occur randomly in the temporal domain.

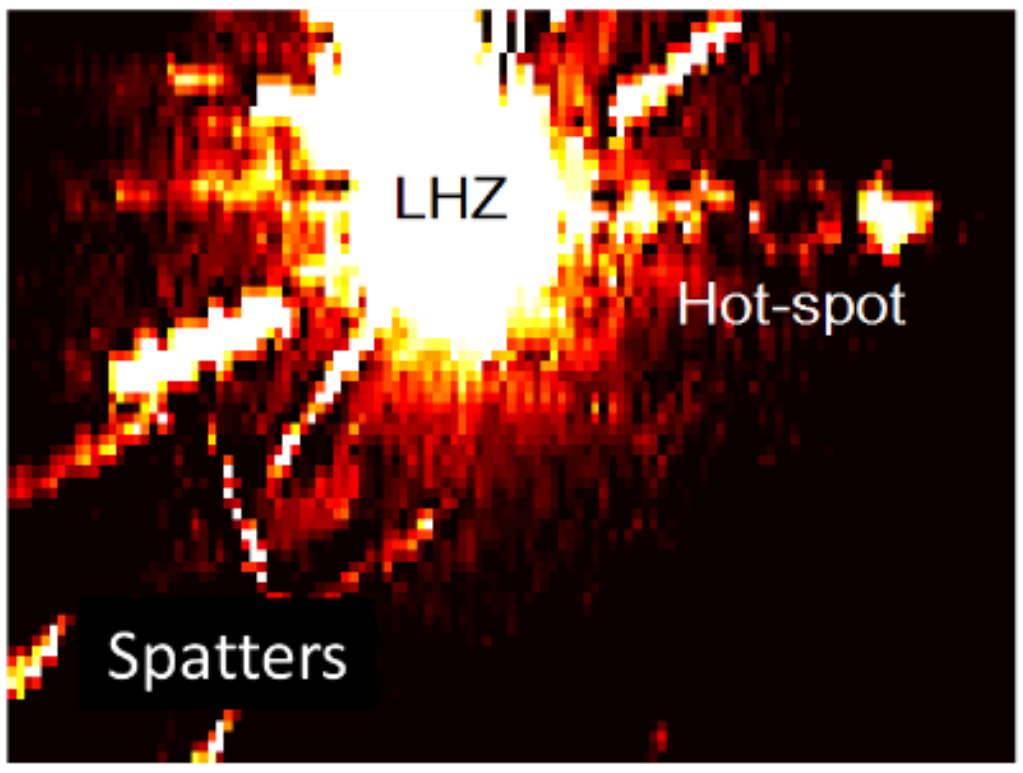

Figure 1. Spatial structure of a single video-frame where a hot-spot is present in addition to the natural process including high-intensity areas corresponding to the LHZ and the spatters produced by the laser-material interaction

Figure 1 shows that the separation of the entire image into the natural process phenomena including the source (i.e., LHZ) and sparse foreground event (i.e., spatters) and an out-of-control anomaly event (i.e., the hot-spot). From Figure 1, we can conclude that using one single image to separate these event is difficult or even impossible. Therefore, the information enclosed by the temporal structure of the video-imaging data should be considered as well. Figure 2 illustrates the temporal behavior of two pixels, i.e., the time-series of their intensity within the recorded video. One pixel was located where the natural process took place, and one is corresponding to the hot-spot. Figure 2 (left panel) shows that the natural spatters and LHZ dynamics yield sudden intensity peaks in the temporal domain, which is representative of the natural temporal pattern of pixel intensities. In the presence of a hot-spot, however, the local heat accumulation causes saturation of the pixel intensity and a slower cooling transitory to the background level. This is representative of the over-heating phenomena that may finally lead to a defective part.

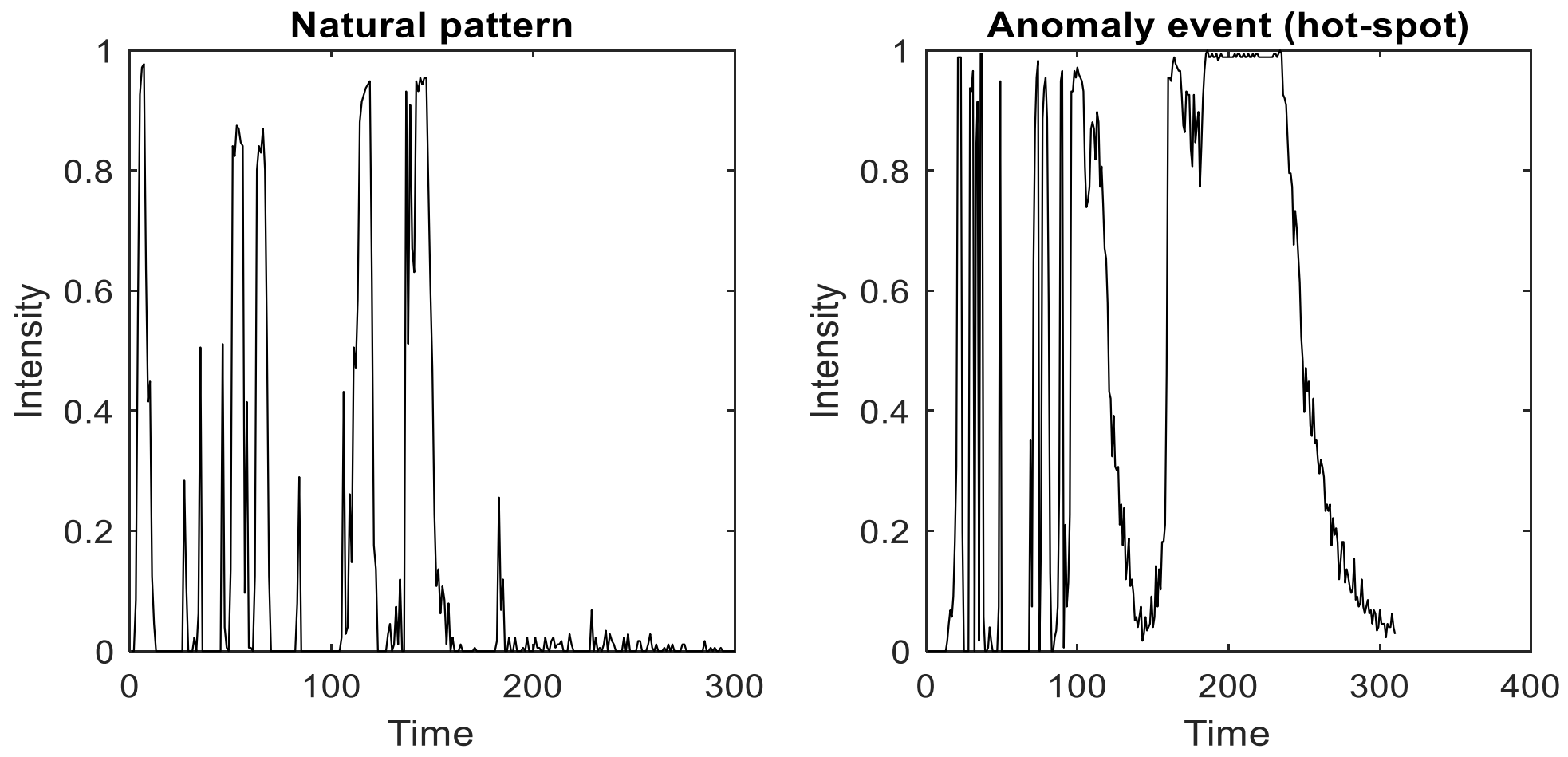

Figure 2. Example of a natural temporal pattern of pixel intensities where peaks are caused by the quick transition of spatters or by the LHZ (left panel); example of an out-of-control temporal pattern of pixel intensities in the presence of a hot-spot (right panel)

The remaining of the paper is organized as follows. Section 2.1 details the proposed penalized spatio-temporal regression model for spatio-temporal data and its recursive estimation procedure. Section 2.2 focuses on the efficient process monitoring techniques by applying LRT on the estimated anomalies. In section 3, a simulation study is used to evaluate the hot-spot detection performances. In section 4, we compare our proposed method against benchmark techniques in the LPBF real case study. Section 5 concludes the paper and presents future directions.

## 2. Literature Review

To address the challenge of detecting localized anomaly events in video-imaging data, three different categories of methods have been proposed in the literature: 1) Principal component analysis (PCA)-based approaches 2) Kernel and basis representation methods 3) Scanning statistics techniques.

The first category includes various PCA and dimension reduction techniques suitable to reduce the dimensions of spatial-temporal data in the framework of statistical process monitoring. For example, Celik (2009) proposed a change detection algorithm for satellite image detection using PCA and K-means clustering. Various multi-variate functional PCA methods and subspace learning methods are developed (Paynabar et al., 2013; Paynabar et al., 2016; Zhang et al., 2018a; Zhang et al., 2018b) to monitor the multi-channel signals.  Yan et al. (2015) compared several famous tensor PCA methods for image-based process monitoring. The major drawback of the PCA-based approach is that although it assumes the low-rank structure of the spatial-temporal dataset, it neglects the locally correlated structure in images. To address this issue, an enhanced method based on a spatially-weighted PCA formulation was proposed by Colosimo and Grasso (2018), which considers the locally correlated structures of the foreground events. However, this method does not fully utilize the sparsity structure of the anomalies, which may cause a detection delay.

The second category includes methods that attempt to model the spatio-temporal structure of an image stream by a set of known spatial or temporal basis, kernel, and covariance structure. To model the smooth spatial or temporal structure of the foreground high-dimensional (HD) data for change point detection, non-parametric techniques such as local kernel regression (Qiu et al., 2010; Zou et al., 2009; Zou et al., 2008) were developed. Gaussian process regression was also proposed for video anomaly detection and representation (Cheng et al., 2015). Spatial-temporal smooth-sparse decomposition (ST-SSD) focuses on detecting sparse anomalies from the smooth spatial and temporal foreground (Yan et al., 2017, 2018). One major disadvantage of ST-SSD is that it assumes that the anomaly at each time is independent. However, in many applications, the anomaly event should be temporally coherent. The temporally coherent and spatially clustered structure of the anomaly are not fully considered in the aforementioned methods.

The third category includes window-based scanning methods developed to deal with anomaly detection in spatial and temporal data (Glaz et al., 2001; Neil et al., 2013). For example, scan statistics use a window-based approach to search the cluster of points in the spatial domain. However, although they are widely used for anomaly detection in scattered point patterns in 3D data, they are not necessarily suitable for spatial-temporal image streams. Other window-based approaches for anomaly detection are developed based on low-dimensional features such as the spatial-temporal gradient and texture information (Li et al., 2014). However, window-based approaches usually require the maximum size of the anomaly to be known in advance, which is not feasible in most applications.

The fourth group (Gobert et al., 2018; Kwon et al., 2018; Okaro et al., 2019; Scime & Beuth, 2018; Sergin & Yan, 2019) of works in the literature focuses on using deep learning methodology for pattern recognition in the spatio-temporal data streams. Techniques such as convolutional neural networks (CNNs) (Krizhevsky et al., 2012) and recurrent neural networks (RNNs) (Mikolov et al., 2010) are commonly used to model complex spatial and temporal structures in image data. However, they typically envisage a supervised learning paradigm and the need for a large and representative training dataset. In the present application, the geometry of the printed slice, together with the underlying dynamics of the melting process, may vary from one layer to another and from one part to another, which could make difficult to collect a labeled dataset that is sufficiently large to implement these methods.

## 3. Proposed Methodology

We first introduce the spatio-temporal decomposition and regression model in Section 3.1. In section 3.2, we discuss how to design process monitoring statistics for monitoring and localization of the hot-spots event. For notation consistency, we use $a$ for a scalar, $\boldsymbol{a}$ for a vector, and $\boldsymbol{A}$ for a matrix.

### 3.1 A Spatio-temporal decomposition model

#### *3.1.1 Formulation*

To capture the spatio-temporal structure of video-imaging data and detect anomaly events, we present the penalized nonparametric regression model and recursive estimation algorithm. We denote $x_{s,t}$ as the intensity value of certain spatio-temporal data (e.g., signal/functional curve or videos) at location $s$ and time $t$. Here, the spatial index $s$ can be either 1D (e.g., signals or functional curves) or 2D (e.g., images). Our proposed model aims to decompose the original image/signal $x_{s,t}$ into the background $\mu_{s,t}$, a natural foreground event $u_{s,t}$, an anomaly foreground event $a_{s,t}$, and noise $e_{s,t}$ as in Eq. (1).

$$x_{s,t} = \mu_{s,t} + u_{s,t} + a_{s,t} + e_{s,t}, \qquad s = 1, \cdots, S, \qquad t = 1, \cdots, T, \tag{1}$$

where we assume that the background $\mu_{s,t}$ is known and $e_{s,t}$ follows an *i.i.d* normal distribution. However, $u_{s,t}$ and $a_{s,t}$ are unknown and should be estimated. We assume that the natural foreground event, $u_{s,t}$, and the anomaly event, $a_{s,t}$, are both sparse in the spatial domain. In addition, the anomaly event is clustered but natural events can be both clustered or scattered.

To model the smooth temporal structure of the anomaly, e.g., a pattern like the one in Figure 2 (right panel), we propose to apply the first-order autoregressive (AR(1)) model with parameter $\theta_s$ to model the temporal behaviour of the anomaly as $a_{s,t} = \theta_s x_{s,t-1}$. Therefore, combining with Eq. (1), we have

$$x_{s,t} = \theta_s x_{s,t-1} + \mu_{s,t} + u_{s,t} + e_{s,t}, \tag{2}$$

where $u_{s,t}$ and $\theta_s$ are the parameters to be estimated. It is worth noting that in the hot-spot detection problem, Eq. (2) has certain physical meaning. For example, during the cooling transitory, the temperature follows $\frac{da_{s,t-1}}{dt} = -\kappa_s x_{s,t-1}$ (Cannon, 1984), where $\kappa_s$ is a constant related to the heat capacity of the material. When the point $s$ exhibits a natural pattern, $\kappa_s \to \infty$, which leads to $a_{s,t} = 0$ or $\theta_s = 0$. However, when the point $s$ exhibits the abnormal pattern discussed above, $\kappa_s$ is small, where $a_{s,t} = (\kappa_s + 1)x_{s,t-1} = \theta_s x_{s,t-1}$, where $\theta_s$ is non-zero. Finally, AR(1) can model a large range of spatio-temporal propagations. For example, first-order models have been widely used in disease propagation in many spatio-temporal studies (Zaman et al., 2008).

The goal is to estimate the foreground event $u_{s,t}$ and the anomaly event $\theta_s$ automatically from the data. We are especially interested in the anomaly event detection, i.e., where $\theta_s \neq 0$. Recall that both $u_{s,t}$ and $\theta_s$ should be sparse. Therefore, at each time $t$, we propose the penalized likelihood $l_t\left(\boldsymbol{\theta}, \{u_{s,t}\}_{s=1,\cdots,S}\right)$ as loss function in Eq (3). Here, we denote $\boldsymbol{\theta}$ as the vector that contains all $\theta_s$, defined as $\boldsymbol{\theta} = [\theta_1, \cdots, \theta_S]^T$.

$$l_t\left(\boldsymbol{\theta}, \{u_{s,t}\}_{s=1,\cdots,S}\right) = \left(\sum_{s,t} \| x_{s,t} - \mu_{s,t} - u_{s,t} - \theta_s x_{s,t-1} \|^2 + \gamma_1 \| u_{s,t} \|_1\right) + \\ + \gamma_2 \| \boldsymbol{\theta} \|_1 + \gamma_3 \| \boldsymbol{\theta} \|_{TV} + \lambda_0 \| \boldsymbol{\theta} \|^2 \tag{3}$$

$\sum_{s,t} \| x_{s,t} - \mu_{s,t} - u_{s,t} - \theta_s x_{s,t-1} \|^2$ is the sum of squared error; The penalty $\| u_{s,t} \|_1$ and $\| \boldsymbol{\theta} \|_1$ leads to the sparse estimation of natural and anomaly events. To encourage the clustered structures of the anomalies, we added the total variance penalty $\| \boldsymbol{\theta} \|_{TV}$, defined as $\| \boldsymbol{\theta} \|_{TV} = \| \boldsymbol{D\theta} \|_1$, which leads to a smooth estimation of the anomalous regions. $\boldsymbol{D}$ is the first-order difference matrix defined as $\boldsymbol{D} = \begin{bmatrix} 1 & -1 & & \\ & \ddots & \ddots & \\ & & 1 & -1 \end{bmatrix}$. To aggregate the loss function $l_t\left(\boldsymbol{\theta}, \{u_{s,t}\}_{s=1,\cdots,S}\right)$ over time, a higher weight can be put on more recent data with the weight decay $\lambda \in (0,1)$ to enable the most up-to-date estimation of the anomalies $\boldsymbol{\theta}$. Here, $l_t$ only depends on $\{u_{s,t}\}$ at time $t$.

$$\min_{\{u_{s,t}\},\boldsymbol{\theta}} L\left(\boldsymbol{\theta}, \{u_{s,t}\}_{s=1,\cdots,S,\ t=1,\cdots,T}\right) = \sum_{t=1}^{T} \lambda^{T-t} l_t\left(\boldsymbol{\theta}, \{u_{s,t}\}_{s=1,\cdots,S}\right) \tag{4}$$

*3.1.2 Recursive Estimation of the Spatio-Temporal Process*

The proposed penalized spatio-temporal regression can effectively model both the temporal and spatial structure of video-imaging data streams. However, since it is required to solve Eq. (4) at each time $t$, an efficient optimization algorithm is needed. In this section, we propose a recursive estimation procedure to update $\theta_s$ and $u_{s,t}$ in a block coordinate manner.

Proposition 1 is proposed to optimize the estimation of the natural event $u_{s,t}$ at time $t$ in Eq. (4):

***Proposition 1.*** Given $\theta_s$ in Eq. (4), $u_{s,t}$ in each time $t$ and each point $s$ can be solved by:

$$u_{s,t} = S(x_{s,t} - \mu_{s,t} - \theta_s x_{s,t-1}, \frac{\gamma_1}{2}), \tag{5}$$

where $S(\cdot,\cdot)$ is the soft thresholding operator defined by $S(x,\gamma) = \text{sgn}(x)\max(x - |\gamma|, 0)$ and $\text{sgn}(x)$ is the sign function. The proof is given in Appendix A.

To optimize the estimation of the anomaly $\boldsymbol{\theta}$ in real-time, we ground on the following proposition:

***Proposition 2.*** Given each $u_{s,t}$ in Eq. (4), $\boldsymbol{\theta}$ can be optimized by

$$\text{argmin}_{\boldsymbol{\theta}}\left(\| \widetilde{\boldsymbol{\Phi}}_{\boldsymbol{T}}\boldsymbol{\theta} - \widetilde{\boldsymbol{\theta}}_{\boldsymbol{T}} \|^2 + \gamma_2 \| \boldsymbol{\theta} \|_1 + \gamma_3 \| \boldsymbol{D\theta} \|_1\right), \tag{6}$$

where $\widetilde{\boldsymbol{\theta}}_T = \left[\tilde{\theta}_{1,T}, \cdots, \tilde{\theta}_{S,T}\right]^T$, $\tilde{\theta}_{s,T} = \frac{\Psi_{s,T}}{\widetilde{\Phi}_{s,T}}$, $\widetilde{\boldsymbol{\Phi}}_{\boldsymbol{T}} = \text{diag}(\widetilde{\Phi}_{1,T}, \cdots, \widetilde{\Phi}_{S,T})$, $\widetilde{\Phi}_{s,T} = \sqrt{\frac{1-\lambda}{1-\lambda^T}\Phi_{s,T} + \lambda_0}$. $\Phi_{s,t}$ and $\Psi_{s,t}$ can be computed recursively as

$$\Phi_{s,t} = \lambda\Phi_{s,t-1} + x_{s,t-1}^2, \Psi_{s,t} = \lambda\Psi_{s,t-1} + x_{s,t-1}(x_{s,t} - \mu_{s,t} - u_{s,t}), \tag{7}$$

with initialization $\Phi_{s,1} = \Psi_{s,1} = 0$.

The proof is given in Appendix B. It is worth noting that all coefficients including $\Phi_{s,t}$, $\Psi_{s,t}$ can be computed recursively in Eq. (7), which leads to a constant updating time. To solve Eq. (6), we follow the Alternating Direction Method of Multipliers (ADMM) algorithm by transforming Eq. (6) to the equivalent problem:

$$\text{argmin}_{\boldsymbol{\theta}}\left(\parallel \widetilde{\boldsymbol{\Phi}}_T\boldsymbol{\theta} - \widetilde{\boldsymbol{\theta}}_T \parallel^2 + \gamma_2 \parallel \boldsymbol{p} \parallel_1 + \gamma_3 \parallel \boldsymbol{q} \parallel_1\right)$$
$$s.t. \boldsymbol{p} = \boldsymbol{D\theta}, \boldsymbol{q} = \boldsymbol{\theta} \tag{8}$$

The augmented Lagrangian for Eq. (8) can be derived in Eq. (9).

$$L(\boldsymbol{\theta}, \boldsymbol{p}, \boldsymbol{q}, \boldsymbol{y}, \boldsymbol{z}) = \parallel \widetilde{\boldsymbol{\Phi}}_T\boldsymbol{\theta} - \widetilde{\boldsymbol{\theta}}_T \parallel^2 + \gamma_2 \parallel \boldsymbol{q} \parallel_1 + \parallel \boldsymbol{p} \parallel_1 - \boldsymbol{y}^T(\boldsymbol{p} - \boldsymbol{D\theta}) - \boldsymbol{z}^T(\boldsymbol{q} - \boldsymbol{\theta}) + \\ + \frac{\rho_p}{2} \parallel \boldsymbol{p} - \boldsymbol{D\theta} \parallel^2 + \frac{\rho_r}{2} \parallel \boldsymbol{q} - \boldsymbol{\theta} \parallel^2, \tag{9}$$

This can be solved via an iterative approach. In the $(k+1)^{th}$ iteration, $\boldsymbol{p}^{(k+1)}, \boldsymbol{q}^{(k+1)}$ and $\boldsymbol{\theta}^{(k+1)}$ can be updated via the following:

$$\boldsymbol{\theta}^{(k+1)} = \boldsymbol{argmin}_{\boldsymbol{\theta}} \parallel \widetilde{\boldsymbol{\Phi}}_T\boldsymbol{\theta} - \widetilde{\boldsymbol{\theta}}_T \parallel^2 + \frac{\rho_p}{2} \parallel \boldsymbol{p}^{(k)} - \boldsymbol{D\theta} \parallel^2 + \frac{\rho_q}{2} \parallel \boldsymbol{q}^{(k)} - \boldsymbol{\theta} \parallel^2 + \\ - {\boldsymbol{y}^{(k)}}^T\left(\boldsymbol{p}^{(k)} - \boldsymbol{D\theta}\right) - {\boldsymbol{z}^{(k)}}^T\left(\boldsymbol{q}^{(k)} - \boldsymbol{\theta}\right)$$
$$\boldsymbol{p}^{(k+1)} = \boldsymbol{argmin}_{\boldsymbol{p}} \frac{\rho_p}{2} \parallel \boldsymbol{p} - \boldsymbol{D\theta}^{(k)} \parallel^2 - {\boldsymbol{y}^{(k)}}^T\left(\boldsymbol{p} - \boldsymbol{D\theta}^{(k)}\right) + \gamma_3 \parallel \boldsymbol{p} \parallel_1 \tag{10}$$
$$\boldsymbol{q}^{(k+1)} = \boldsymbol{argmin}_{\boldsymbol{q}} \frac{\rho_q}{2} \parallel \boldsymbol{q} - \boldsymbol{\theta}^{(k)} \parallel^2 - {\boldsymbol{z}^{(k)}}^T\left(\boldsymbol{q} - \boldsymbol{\theta}^{(k)}\right) + \gamma_2 \parallel \boldsymbol{q} \parallel_1$$

And the dual variable $\boldsymbol{y}^{(k+1)}, \boldsymbol{z}^{(k+1)}$ should be updated via:

$$\boldsymbol{y}^{(k+1)} = \boldsymbol{y}^{(k)} - \rho_p\left(\boldsymbol{p}^{(k)} - \boldsymbol{D\theta}^{(k)}\right) \tag{11}$$

$$\boldsymbol{z}^{(k+1)} = \boldsymbol{z}^{(k)} - \rho_q(\boldsymbol{q}^{(k)} - \boldsymbol{\theta}^{(k)}) \tag{12}$$

$\boldsymbol{\theta}^{(k+1)}$ can be updated via:

$$\boldsymbol{\theta}^{(k+1)} = \left(2\widetilde{\boldsymbol{\Phi}}_T^2 + \rho_q\boldsymbol{I} + \rho_p\boldsymbol{D}^T\boldsymbol{D}\right)^{-1}\left(2\widetilde{\boldsymbol{\Phi}}_T\widetilde{\boldsymbol{\theta}}_T + \rho_p\boldsymbol{D}^T\boldsymbol{p}^{(k)} + \rho_q\boldsymbol{q}^{(k)} - \boldsymbol{D}\boldsymbol{y}^{(k)} - \boldsymbol{z}^{(k)}\right) \tag{13}$$

It is worth noting that computing $\boldsymbol{\theta}^{(k+1)}$ directly in Eq. (13) may not be computationally efficient due to the matrix inverse operator. However, the matrix inverse $\left(2\widetilde{\boldsymbol{\Phi}}_T^2 + \rho_q I + \rho_p\boldsymbol{D}^T\boldsymbol{D}\right)^{-1}$ can be computed before all iterations to save time. Furthermore, it is possible to demonstrate that $\boldsymbol{p}^{(k+1)}$ and $\boldsymbol{q}^{(k+1)}$ can be solved efficiently by soft-thresholding via:

$$\boldsymbol{p}^{(k+1)} = S\left(\boldsymbol{D}\boldsymbol{\theta}^{(k)} + \frac{1}{\rho_p}\,\boldsymbol{y}^{(k)}, \frac{\gamma_3}{\rho_p}\right) \tag{14}$$

$$\boldsymbol{q}^{(k+1)} = S\left(\boldsymbol{\theta}^{(k)} + \frac{1}{\rho_q}\,\boldsymbol{z}^{(k)}, \frac{\gamma_2}{\rho_q}\right) \tag{15}$$

Here $\rho_p, \rho_q$ are fixed constants. We update $\rho_p$ as $\rho_p = \gamma\rho_p$ only if $\| \boldsymbol{p}^{(k+1)} - \boldsymbol{D}\boldsymbol{\theta}^{(k+1)} \|_2 \geq \alpha \| \boldsymbol{p}^{(k)} - \boldsymbol{D}\boldsymbol{\theta}^{(k)} \|_2$. A similar rule can be applied to $\rho_q$, which has shown to result in better computation efficiency (Chan et al., 2011).

---

**Algorithm 1**: Recursive algorithm for penalzied spatial-temporal regression

---

For each time $\boldsymbol{t}$:

Estimate $\boldsymbol{\Phi}_{s,t} = \boldsymbol{\lambda}\boldsymbol{\Phi}_{s,t-1} + \boldsymbol{x}_{s,t-1}^{2}, \boldsymbol{\Psi}_{s,t}^{(0)} = \boldsymbol{\lambda}\boldsymbol{\Psi}_{s,t-1}^{(K)} + \boldsymbol{x}_{s,t-1}(\boldsymbol{x}_{s,t} - \boldsymbol{\mu}_{s,t})$

While not converged.

For each $\boldsymbol{s}$

$\boldsymbol{\Psi}_{s,t}^{(k+1)} = \boldsymbol{\Psi}_{s,t}^{(0)} - \boldsymbol{x}_{s,t-1}\boldsymbol{u}_{s,t}^{(k+1)}$

$\boldsymbol{u}_{s,t}^{(k+1)} = \boldsymbol{S}(\boldsymbol{x}_{s,t} - \boldsymbol{\mu}_{s,t} - \boldsymbol{\theta}_{s}^{(k)}\boldsymbol{x}_{s,t-1}, \frac{\boldsymbol{\gamma}_1}{\boldsymbol{2}})$

End

$$\boldsymbol{\theta}^{(k+1)} = \left(2\tilde{\boldsymbol{\Phi}}_T^2 + \boldsymbol{\rho}_q\boldsymbol{I} + \boldsymbol{\rho}_p\boldsymbol{D}^T\boldsymbol{D}\right)^{-1}\left(2\tilde{\boldsymbol{\Phi}}_T\tilde{\boldsymbol{\theta}}_T + \boldsymbol{\rho}_p\boldsymbol{D}^T\boldsymbol{p}^{(k)} + \boldsymbol{\rho}_q\boldsymbol{q}^{(k)} - \boldsymbol{D}\boldsymbol{y}^{(k)} - \boldsymbol{z}^{(k)}\right)$$

$$\boldsymbol{p}^{(k+1)} = \boldsymbol{S}\left(\boldsymbol{D}\boldsymbol{\theta}^{(k)} + \frac{\boldsymbol{1}}{\boldsymbol{\rho}_p}\,\boldsymbol{y}^{(k)}, \frac{\boldsymbol{\gamma}_3}{\boldsymbol{\rho}_p}\right)$$

$$\boldsymbol{q}^{(k+1)} = \boldsymbol{S}\left(\boldsymbol{\theta}^{(k)} + \frac{\boldsymbol{1}}{\boldsymbol{\rho}_q}\,\boldsymbol{z}^{(k)}, \frac{\boldsymbol{\gamma}_2}{\boldsymbol{\rho}_q}\right)$$

End

End

---

*3.1.3 Efficient Computation for 2D Image*

Even though all operators in Algorithm 1 have a closed-form solution and can be computed analytically in each iteration, it is possible to further speed up the computation for the real-time implementation of the proposed approach. This section presents a few approximations that can be applied to tackle this issue. One approximation to solve Eq. (13) consists of approximating $\widetilde{\boldsymbol{\Phi}}_T^2$ with an identity matrix $\widetilde{\boldsymbol{\Phi}}_T^2 \approx \lambda_0 \boldsymbol{I}$. Here we denote the solution of $\boldsymbol{\theta}$ for given $\gamma_2$ and $\gamma_3$ in Eq. (6) as $\widehat{\boldsymbol{\theta}}_{\gamma_2,\gamma_3}$

Under this approximation, it is possible to prove the following proposition (Xin et al., 2014). It is worth noting that in case this approximation is not precise, we can still use Algorithm 1 for the estimation of $\widetilde{\Phi}_{s,T}$ without using this approximation.

***Proposition 3.*** Given $\gamma_3 = 0$ and $\widetilde{\boldsymbol{\Phi}}_T^2 = \lambda_0 \boldsymbol{I}$ in Eq. (13), $\theta_s$ can be solved via soft-thresholding in Eq. (16).

$$\widehat{\boldsymbol{\theta}}_{\gamma_2,\gamma_3} = \boldsymbol{S}(\widehat{\boldsymbol{\theta}}_{\boldsymbol{0},\gamma_3}, \boldsymbol{\gamma_2}) \quad (16)$$

The proof of Proposition 3 can be seen in (J. Liu et al., 2010). Based on Proposition 3, we only need to solve a special case where $\gamma_2 = 0$ as $\widehat{\boldsymbol{\theta}}_{0,\gamma_3}$. Another advantage is that Proposition 3 allows us to solve $\widehat{\boldsymbol{\theta}}_{\gamma_2,\gamma_3}$ with multiple $\gamma_2$ efficiently, which can be used to better design the testing statistics as mentioned in Section 5.1.

Another benefit of this approximation is that $\left(\boldsymbol{2}\lambda_0\boldsymbol{I} + \rho_q\boldsymbol{I} + \rho_p\boldsymbol{D}^T\boldsymbol{D}\right)^{-\boldsymbol{1}}$ can be computed efficiently. The detailed procedure is as follow: since $\boldsymbol{D}^{\boldsymbol{T}}\boldsymbol{D}$ are block-circulant, it can efficiently be computed or diagonalized by 2D discrete Fourier transform (DFT) $\boldsymbol{F}$ as $\boldsymbol{D}^{\boldsymbol{T}}\boldsymbol{D} = \boldsymbol{F}^{\boldsymbol{T}}\boldsymbol{\Lambda}\boldsymbol{F}$ and $\boldsymbol{F}$ is the discrete Cosine transformation matrix and can be computed efficiently via the discrete Cosine transformation with complexity $O(n \log n)$ instead of complexity $O(n^3)$. For more details of diagonalizing a circulant matrix using DFT, refer to Theorem 3.2.2 in (Davis, 2012). In the end, $\left((2\lambda_0+\rho_q)\boldsymbol{I}+\rho_p\boldsymbol{D}^T\boldsymbol{D}\right)^{-\mathbf{1}} = \boldsymbol{F}^{\boldsymbol{T}}\left((2\lambda_0+\rho_q)\boldsymbol{I}+\rho_p\boldsymbol{\Lambda}\right)^{-\mathbf{1}}\boldsymbol{F}$.

The proposed method with the approximations for computational efficiency improvement can be generalized to 2D images. Since an image can be represented by a matrix, we need to introduce two indices $(x,y)$ to denote the location of each pixel $s$. Here, we denote $\boldsymbol{\Theta}$ as the coefficient matrix, where $\boldsymbol{\Theta}(x,y)$ is the value of the pixel $(x,y)$. Another major difference is the definition of the total variance penalty, which should be defined in both $x$ and $y$ directions as $\| \boldsymbol{\theta} \|_{TV} = \| \boldsymbol{D_x}\boldsymbol{\theta} \|_1 + \| \boldsymbol{D_y}\boldsymbol{\theta} \|_1$. $\boldsymbol{D_x}$ and $\boldsymbol{D_y}$ are the first-order finite-difference operators along the horizontal and vertical direction, as $\boldsymbol{D_x}\boldsymbol{\theta} = \mathbf{vec}(\boldsymbol{\Theta}(x+1,y) - \boldsymbol{\Theta}(x,y))$ and $\boldsymbol{D_y}\boldsymbol{\theta} = \mathbf{vec}(\boldsymbol{\Theta}(x,y+1) - \boldsymbol{\Theta}(x,y))$. Furthermore, we can define a matrix $\boldsymbol{D} = \left[\boldsymbol{D}_x^T\ \boldsymbol{D}_y^T\right]^T$ such as $\| \boldsymbol{D}\boldsymbol{\theta} \|_1 = \| \boldsymbol{D_x}\boldsymbol{\theta} \|_1 + \| \boldsymbol{D_y}\boldsymbol{\theta} \|_1$. Correspondingly, if we define $\boldsymbol{p} = [\boldsymbol{p}_x^T\ \ \boldsymbol{p}_y^T]$ and $\boldsymbol{y} = [\boldsymbol{y}_x^T\ \ \boldsymbol{y}_y^T]$, the exact updating rule in Algorithm 1 can be used.

### 3.2 Proposed Process Monitoring based on Penalized Spatio-Temporal Regression

In the context of statistical process control (SPC), process monitoring includes two stages, namely Phase I and Phase II (Oakland, 2007). The monitoring statistic that can be used for Phase I and Phase II analysis is described in Section 2.2.1. Phase I analysis can be used to estimate the tuning parameters and the control limit. The detailed procedure is discussed in Section 2.2.2. Finally, the localization of the hotspots is discussed in section 2.2.3.

#### *3.2.1 Monitoring Statistics*

In this section, we describe a statistical process monitoring approach for video-imaging data that combines our proposed penalized spatial-temporal regression with a sequential LRT. Furthermore, we also discuss how the proposed method can be used for locating the anomaly after a change is detected. As we discussed before, the goal is to detect and monitor the occurrence of the clustered anomaly event presented in the spatio-temporal dataset. We are interested in signaling an event occurring in the image location $s$ where $\theta_s \neq 0$. Therefore, we formulate the monitoring problem as a sequential hypothesis testing problem with the null hypothesis that no anomaly event is happening.

The rate of each point can be estimated individually as $\widehat{\boldsymbol{\theta}}_{\mathbf{0,0}}\,(t)$ with $\gamma_2 = \gamma_3 = 0$. Since we assume that an anomaly event covers just a small portion of the entire image, our proposed estimator with $L_1$ and total variance penalty is used to accurately represent the sparse clustered structure of the anomaly. This provides an accurate estimation of the anomalous event by $\widehat{\boldsymbol{\theta}}_{\gamma_2,\gamma_3}(t)$ at each time $t$ with tuning parameters $\gamma_2, \gamma_3$. Therefore, at each time $t$, we perform the following hypothesis test:

$$H_0: \boldsymbol{\theta} = 0 \qquad H_1: \boldsymbol{\theta} = \delta \widehat{\boldsymbol{\theta}}_{\gamma_2,\gamma_3}(t)$$

Following the procedure in (Yan et al., 2018; Zou & Qiu, 2009), we can derive the testing statistic in Eq. (17).

$$\tilde{T}_{\gamma_2,\gamma_3}(t) = \frac{\left(\left(\widehat{\boldsymbol{\theta}}_{\gamma_2,\gamma_3}(t)\right) \;\; \boldsymbol{\theta}_{0,0}\,(t)\right)^2}{\| \, \widehat{\boldsymbol{\theta}}_{\gamma_2,\gamma_3}(t) \, \|^2} \tag{17}$$

Before $\tilde{\boldsymbol{T}}_{\gamma_2,\gamma_3}$ can be used for process monitoring, the regularization $\gamma_2, \gamma_3$ should be chosen carefully since it plays an important role to control the sparsity and smoothness of $\widehat{\boldsymbol{\theta}}_{\gamma_2,\gamma_3}$. Therefore, to make the testing statistics robust for the tuning parameter selections, the modified testing statistic is defined as the tuning parameters that achieve the best normalized OC signal:

$$\tilde{T}(t) = \max_{(\gamma_1,\gamma_2,\gamma_3)\in\Gamma} \frac{\tilde{T}_{\gamma_2,\gamma_3}(t) - E\left(\tilde{T}_{\gamma_2,\gamma_3}\right)}{\sqrt{Var(\tilde{T}_{\gamma_2,\gamma_3})}} \tag{18}$$

Here, the mean and variance of the $\tilde{T}_{\gamma_2,\gamma_3}$ can be estimated by the sample mean and sample variance of $\tilde{T}_{\gamma_2,\gamma_3}$ from the IC data. Finally, we choose a control limit $L > 0$ for Eq. (8) and if $\tilde{T}(t) > L$, the monitoring scheme would trigger an out-of-control (OOC) alarm at time $\boldsymbol{t}$. Let $\boldsymbol{\Gamma}$ be the set of parameters $(\gamma_1, \gamma_2, \gamma_3)$. The selection of the tuning parameter, the control limit $\boldsymbol{L}$, and other parameters is discussed in Section 3.2.

*3.2.2 Tuning parameter selection*

In this section, we discuss how to select tuning parameters $\gamma_1, \lambda_0, \Gamma, \lambda$, and the control limit $L$. A larger $\gamma_1$ will lead to a sparser estimation of the sparse event. Selecting a non-zero $\lambda_0$ is helpful for the case of large colinearity. Therefore, we can select $\lambda_0$ such that the detected natural foreground event takes roughly about 5% of all the pixels. The set of parameters $(\gamma_1, \gamma_2, \gamma_3)$ can be selected as $\Gamma = \left\{\left(\frac{\gamma_{1max}}{n_{\gamma_1}} j_1, \frac{\gamma_{2max}}{n_{\gamma_2}} j_2, \frac{\gamma_{3max}}{n_{\gamma_3}} j_3\right), j_1 = 0, \cdots, n_{\gamma_1}, j_2 = 0, \cdots, n_{\gamma_2}, j_3 = 0, \cdots, n_{\gamma_3}\right\}$, where $\gamma_{2max}$ can be selected as the $\gamma_2$ value such that $\widehat{\boldsymbol{\theta}}_{\gamma_2,0} = 0$ and $\gamma_{3max}$ can be selected as the $\gamma_3$ value such that $\widehat{\boldsymbol{\theta}}_{0,\gamma_3} = c$. Selecting a larger $n_{\gamma_2}$ and $n_{\gamma_3}$ may lead to a better detection power, but it yields a higher computational effort. In our experiments, we selected $n_{\gamma_1} = n_{\gamma_2} = 5$ and $n_{\gamma_3} = 2$ to balance the computational effort and detection power. The choice of $\lambda$ also depends on the different applications and the duration and intensity of the true mean shift. A larger $\lambda$ improves the detection of a smaller meanshift but it inflates the mean time to detect a larger meanshift. The choice of $\lambda$ can follow the same procedure as the traditional exponentially-weighted moving average (EWMA) control chart (Lu & Reynolds Jr, 1999). Additional details about how $\lambda$ would affect the detection power are discussed in the simulation study. Finally, we can choose the control limit $L > 0$ to achieve a given IC average run length (ARL). Some numerical searching algorithms, such as bisection search, can be applied (Zhang et al., 2018b).

*3.2.3 Localization of Detected Changes*

After the proposed control chart triggers an OOC signal, the next step consists of identifying where the anomalous event has occurred. This information would help the process engineers identify and later eliminate the potential root causes. Suppose the control chart triggers a signal at time $\tau$, we will first find the best pair of $(\gamma_1, \gamma_2, \gamma_3)$, which optimizes Eq. (17). Finally, the non-zero elements of the corresponding $\widehat{\boldsymbol{\theta}}_{\gamma_2,\gamma_3}$ can be used to identify the location of the anomaly event.

## 4. Simulation Analysis

In this section, we will use the simulation analysis to evaluate the performance of the proposed algorithm, which carried out by artificially injecting hot-spot events in different locations and with different sizes into a real video-image stream acquired during an in-control LPBF process. More details of the simulation setup are discussed in Section 3.1. We will then discuss the performance evaluation of the proposed method and the benchmark methods in Section 3.2. Finally, we perform a sensitivity analysis of how the tuning parameters would affect the accuracy in Section 3.3.

### 3.1 Simulation Setup

The cylindrical shape of diameter 16 mm was produced via LPBF of AISI 316L powder (average particle size of about 25-30 μm) on Renishaw® AM250 system. The post-process inspections of the as-built part allowed judging the process as in-control and the part as defect-free. A 150 fps video-sequence was acquired during the realization of one layer of the part by using the setup shown in Figure 3, which consists of an OlympusTM I-speed 3 camera (CMOS sensor) mounting a 50 mm lens placed outside the build chamber viewport. Additional details about the experimental setup and the LPBF process are discussed in (Colosimo & Grasso, 2018).

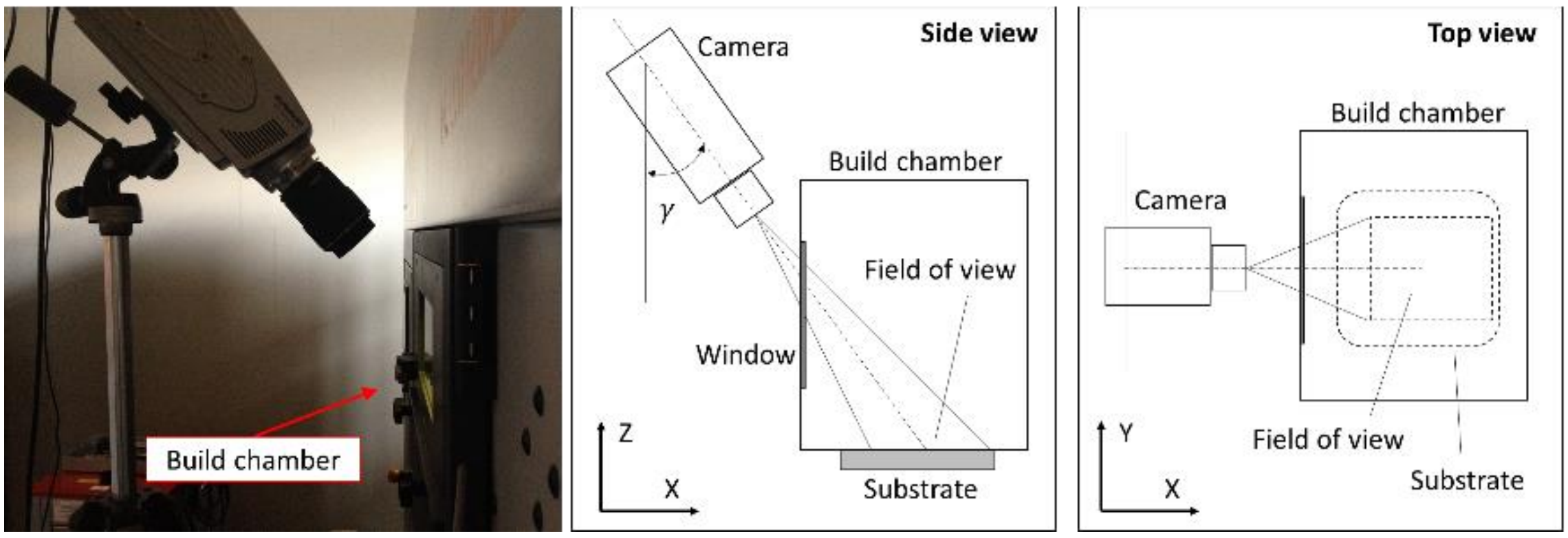

Figure 3 – Experimental setup used for high-speed video imaging with OlympusTM I-speed 3 camera placed outside a commercial Renishaw® AM250 system

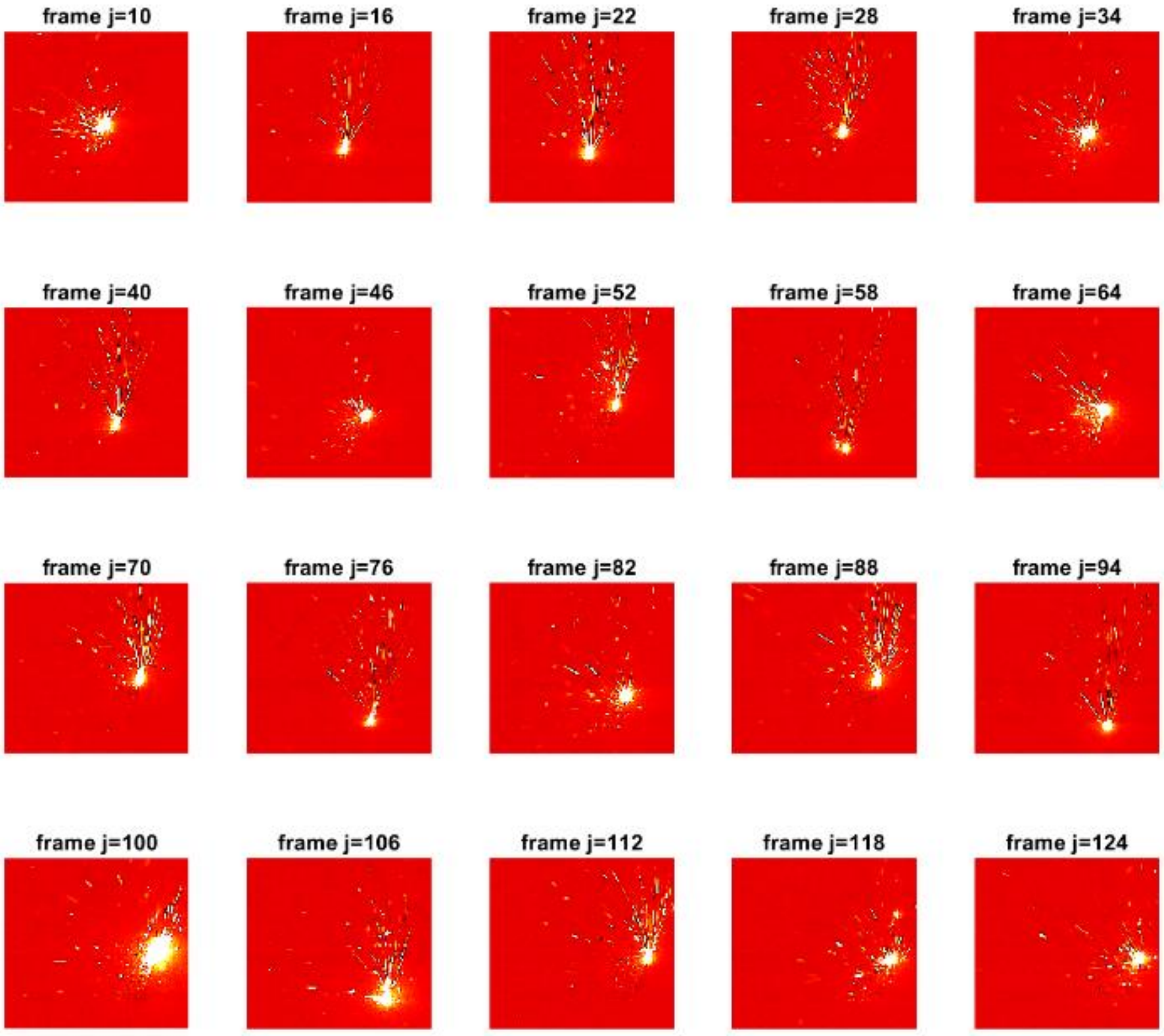

Figure 4. Examples of video frames (in false colors) acquired during the LPBF of a cylindrical part used as a reference for the generation of simulated hot-spot events

As an example, Figure 4 shows a subset of the original video frames, where a circular slice of the vertical cylinder was produced. The frame size was 126 x 136 pixels.

Based on this video-sequence, hot-spot events were simulated by modifying the pixel intensity of a number $n$ of adjacent pixels for a given duration, $\tau$. Five levels of hot-spot size were considered, corresponding to $n = [4, 9\ 20, 45, 80]$. To make the simulated hot-spot as realistic as possible, the following procedure was applied. Let $(x_i, y_i)_{LHZ}$ be the location of the centroid of the LHZ identified in the $i$-th frame. The LHZ was identified as the largest connected component in the image in accordance with previous studies (Repossini et al., 2017). Then, the hot-spot was injected starting from the $i$-th frame and it was visible in a fixed location for the next $\tau$ frames, with its centroid placed in $(x_i, y_i)_{LHZ}$, while the LHZ displaces along the predefined scan path. Indeed, real hot-spots originate just after the laser scanning of a region of the slice.

An example of one video frame where simulated hot-spots were injected with different sizes is shown in Figure 5. In this case, a cross-shaped hot-spot was simulated to simplify the visual identification of the anomaly in the video frames.

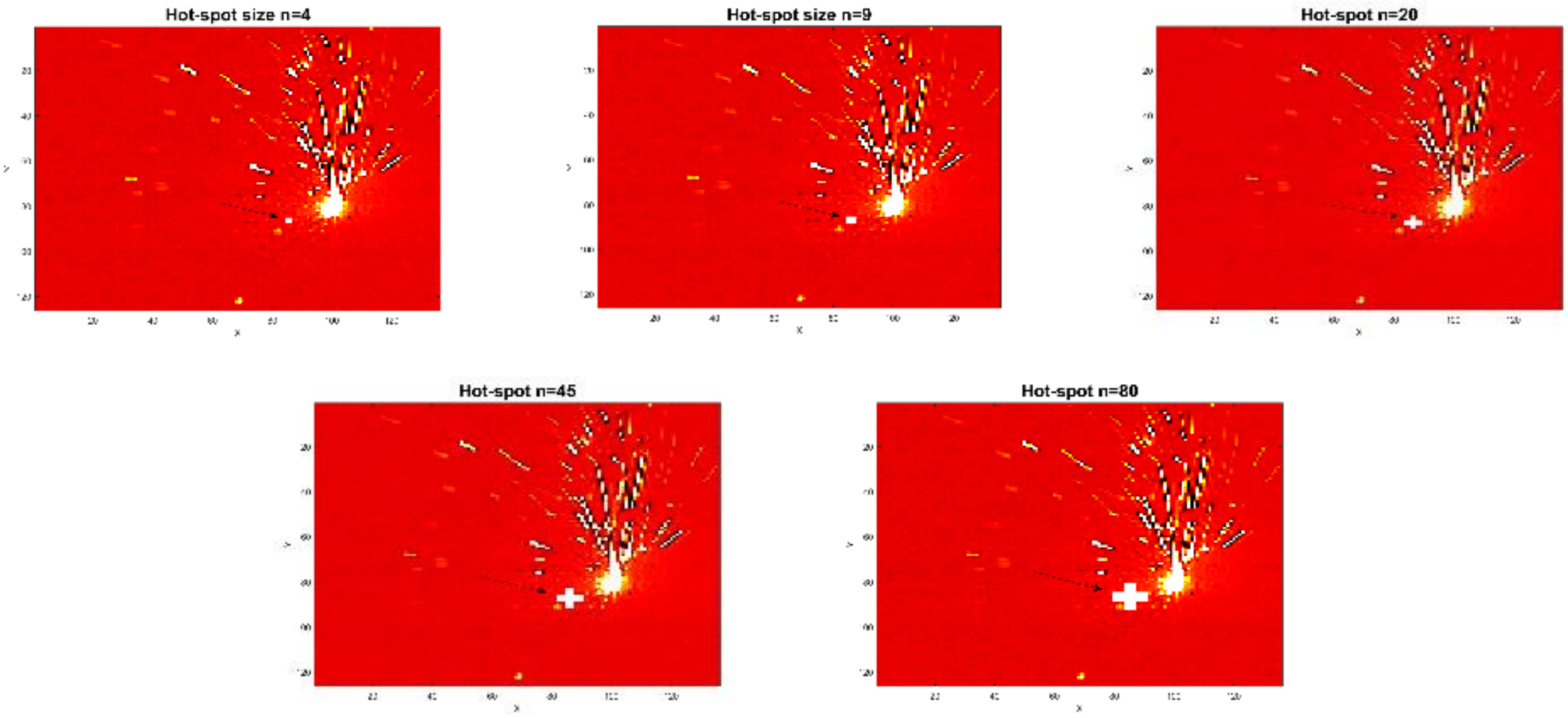

Figure 5. Examples of one original video frame (false colors) with an injection of simulated hot-spot of different sizes

The simulated hot-spot consists of a saturated intensity* ($a_{s,t} = 255$) for several consecutive frames followed by a slow cooling transitory (i.e., a pixel intensity decrease) to the average background

* In 8-bit images, the original pixel intensity ranges between 0 and 255.

intensity. A sigmoid function was used to generate this pattern, accordingly to the following expression:

$$a_{s,t} = \frac{255}{1 + \exp(0.2(t - H\tau))}, \qquad t = 1, \dots, \tau, \tag{19}$$

where $H$ is a constant that controls the shape of the cooling profile in the hot-spot regions. Eq. (19) with $H = 0.95$ was used in this study to generate a realistic hot-spot pattern over time. We refer the reader to (Colosimo & Grasso, 2018) for additional details. The hot-spot simulation was repeated for 100 different locations.

### 3.2 Performance Evaluation

We compared the proposed method against four benchmark methods available from the literature. The first benchmark approach is Hoteling's $T^2$ control chart (Hotelling, 1992) (denoted as "T2"), applied to the vectorized frames of the video, where each pixel is treated as a variable and each frame as a new observation. This is representative of a basic control charting scheme that can be applied to video-images by practitioners. The second benchmark method is a PCA-based control chart applied to vectorized images (Nomikos & MacGregor, 1995). This is representative of the S-mode PCA-based approach mentioned in (Colosimo & Grasso, 2018) and it is representative of the basic way to apply the PCA to video-imaging data  (denoted as "PCA"). The third benchmark method consists of a Lasso-based control chart (denoted as "LASSO"). In this case, we implemented the procedure proposed by (Zou & Qiu, 2009), in which LASSO (Tibshirani, 1996) is used to first identify the sparse change direction and the likelihood ratio test is applied for change detection and anomaly identification.  The forth benchmark method is based on the spatio-temporal smooth sparse decomposition (denoted as "STSSD") to separate the anomaly event from the background events (Yan et al., 2018). The last benchmark method is the spatially weighted T-mode PCA (denoted as "ST-PCA"), proposed by Colosimo and Grasso (2018). This is representative of the current state-of-the-art methodologies for hot-spot detection in LBPF via in-situ video-imaging. In the T-mode PCA formulation (Jolliffe, 2002; Tsutsumida et al., 2017), the video frames are treated as variables and image pixels as observations. This allows one to capture the temporal auto-correlation of pixel intensities over consecutive frames. The underlying idea of the spatially weighted T-mode PCA consists of incorporating the pixel spatial correlation into the projection operation entailed by the T-mode PCA. The resulting ST-PCA was combined with a recursive updating scheme to iterative include new video frames for in-line hot-spot detection. A k-means clustering-based alarm rule was eventually proposed to signal an alarm in the presence of a region of the image where the hot-spot event occurred. For T2, PCA and LASSO, since the algorithms cannot handle the temporal – dependent dynamics of the LHZ, they rely on a pre-processing step that consists of removing the largest connected component corresponding to the LHZ from each frame. For all the methods, we selected the control limit based on the IC samples with the false positive rate of 0.01.

The performances of compared methods were estimated by means of different metrics. The average run length (ARL) was used to quantify how fast each method can detect the hot-spot. Precision and recall scores were used to quantify the classification accuracy when the defect is detected (Powers, 2011). The precision score can be defined as the ratio between the pixels belonging to the hot-spot region that were correctly detected by the monitoring method and the overall number of detected pixels. A precision score equal to 1 means that the monitoring method detected only hot-spot pixels, and hence no false alarm outside the hot-spot region is produced. The recall score can be defined as the ratio between the pixels belonging to the hot-spot region that were correctly detected by the monitoring method and the overall number of pixels belonging to the hot-spot. A recall score equal to 1 means that the monitoring method can detect the entire hot-spot region, and hence, no false negative is present. One additional metric was considered: it is the F1 score, which is the harmonic mean of the precision and recall score (Rijsbergen, 1979).

Table 1 summarizes the comparison between the proposed approach and all other competing methods in the presence of simulated hot-spots of different sizes.

Table 1 Average values of performances indexes for different methods and different hot-spot sizes (the standard deviation of mean values is reported in parentheses)

| Hot-spot size | Method | ARL | Precision | Recall | F1 |
|---|---|---|---|---|---|
| Small | Proposed | **3.39 (1.57)** | 0.88 (0.27) | 0.98 (0.14) | 0.90 (0.24) |
| | T-square | 75.31 (47.01) | 0.01 (0.01) | 0.33 (0.47) | 0.01 (0.02) |
| (n=4) | PCA | 87.66 (49.48) | 0.00 (0.01) | 0.30 (0.43) | 0.01 (0.01) |
| | Lasso | 90.51 (49.12) | 0.01 (0.01) | 0.23 (0.42) | 0.00 (0.00) |
| | STSSD | 77.28 (53.90) | 0.00(0.00) | 0.35 (0.47) | 0.01(0.01) |
| | STPCA | 73.19 (1.57) | **1.00 (0)** | **1.00 (0)** | **1.00 (0)** |
| Med-small (n=9) | Proposed | **2.65 (1.14)** | 0.91 (0.19) | 0.98 (0.14) | 0.94 (0.17) |
| | T-square | 75.31 (47.01) | 0.01 (0.02) | 0.34 (0.47) | 0.02 (0.03) |
| | PCA | 81.96 (49.16) | 0.01 (0.02) | 0.35 (0.45) | 0.02 (0.03) |
| | Lasso | 84.81 (49.49) | 0.01 (0.02) | 0.41 (0.46) | 0.01 (0.00) |
| | STSSD | 72.74 (54.94) | 0.00(0.00) | 0.46 (0.50) | 0.01(0.01) |
| | STPCA | 68.33 (2.10) | **1.00 (0)** | **1.00 (0)** | **1.00 (0)** |
| Medium (n=20) | Proposed | **2.29 (0.78)** | **0.97 (0.11)** | 0.99 (0.10) | **0.98 (0.10)** |
| | T-square | 81.01 (46.65) | 0.02 (0.04) | 0.29 (0.45) | 0.04 (0.07) |
| | PCA | 84.81 (47.61) | 0.02 (0.03) | 0.31 (0.43) | 0.04 (0.06) |
| | Lasso | 85.76 (47.75) | 0.02 (0.04) | 0.52 (0.45) | 0.04 (0.08) |
| | STSSD | 59.62 (57.87) | 0.01(0.00) | 0.60 (0.49) | 0.02 (0.00) |
| | STPCA | 63.08 (2.27) | 0.94 (0.01) | **1.00 (0)** | 0.97 (0.01) |
| Med-large (n=45) | Proposed | **2.01 (0.52)** | **0.98 (0.10)** | 0.99 (0.10) | **0.98 (0.10)** |
| | T-square | 80.21 (49.48) | 0.05 (0.08) | 0.32 (0.47) | 0.09 (0.13) |
| | PCA | 83.00 (48.66) | 0.05 (0.07) | 0.32 (0.44) | 0.08 (0.12) |
| | Lasso | 54.56 (46.83) | 0.14 (0.12) | 0.74 (0.37) | 0.22 (0.19) |
| | STSSD | 37.09 (47.73) | 0.02(0.00) | 0.81 (0.39) | 0.04 (0.01) |
| | STPCA | 58.39 (2.43) | 0.88 (0.01) | **0.99 (0.003)** | 0.93 (0.01) |
| Large | Proposed | **1.20 (0.58)** | **0.87 (0.16)** | **0.99 (0.10)** | **0.92 (0.13)** |
| | T-square | 65.03 (50.77) | 0.14 (0.14) | 0.50 (0.50) | 0.22 (0.22) |
| (n=80) | PCA | 74.50 (53.21) | 0.12 (0.14) | 0.46 (0.47) | 0.19 (0.21) |
| | Lasso | 36.73 (45.28) | 0.33 (0.20) | 0.75 (0.43) | 0.45 (0.27) |
| | STSSD | 45.75 (53.82) | 0.04(0.00) | 0.61 (0.42) | 0.07 (0.02) |
| | STPCA | 65.76 (2.07) | 0.83 (0.01) | 0.96 (0.01) | 0.89 (0.01) |

Table 1 shows that the proposed method is much faster than all other methods in detecting the hot-spot. In terms of ARL, the proposed method can detect the hot-spot in less than 3-4 frames from the hot-spot event injection for all simulated sizes. On the contrary, no other method was able to signal the hot-spot in less than 50 – 60 frames since its onset. Indeed, competitor methods rely on the detection of an out-of-control shift in the temporal and/or spatial auto-correlation of pixel intensities, which requires a sufficient number of frames before an alarm can be signaled. The proposed approach, instead, entails a model of both the natural and out-of-control patterns that may arise in the video-imaging data. Therefore, the hot-spot event can be detected since its onset stage because its occurrence yields a sudden shift of the corresponding parameter in the spatio-temporal model.

In terms of hot-spot localization accuracy, the Hotelling's $T^2$ control chart, the basic PCA-based approach, the lasso-based control chart, and spatio-temporal smooth sparse decomposition are not only slower than the proposed approach in detecting the hot-spot, but they are also less accurate. The precision score of these three competitors is always very low, which means that the signal as out-of-control a large portion of the video frame. This makes these methods not effective in the present application. The $T^2$ control chart and the PCA-based control chart fail because they focus on detecting a global variation of video-imaging data patterns, whereas the hot-spot event is local in nature, with a reduced effect on both the average pixel intensity and the global variability. Unlike them, the lasso-based control chart allows dealing with the spatial structure of video frames, and hence, it is potentially able to detect local events. However, as discussed in Section 1, the spatial information may not be sufficient to distinguish the hot-spot event from other natural foreground events, i.e., the LHZ and the spatters generated by the laser-material interaction. Indeed, the lasso-based control chart fails because it does not consider any temporal structure of the video-imaging data. The performance of STSSD falls behind the proposed method dramatically. STSSD assumes that the background is smooth and anomaly is abrupt change, which violates our assumption that the background is random but the anomaly is spatially clustered and temporally consistent. The ST-PCA methodology yields comparable and accurate results in localizing the hot-spot, but it requires a sufficient number of video frames to properly identify an anomalous auto-correlation pattern in the pixel intensities within the hot-spot region. Moreover, in the presence of the largest simulated hot-spot event, the performances of the ST-PCA based methodology were slightly worse than those for smaller hot-spots. Indeed, when the hot-spot becomes larger, the overlap between the hot-spot itself and the LHZ increases, reducing the capability of distinguishing the two regions into two separate clusters. This issue does not affect the proposed method, whose performance improves as the hot-spot size increases.

Figure 6 shows the detected anomalies for different methods in one simulation run. In that run, only the proposed method and the ST-PCA signaled an alarm. Figure 6 (left panels) shows original images corresponding to the video frames where an alarm was signaled. Figure 6 (right panels) shows the regions of the frames (white) signaled as a detected anomaly. In Figure 6, both the proposed method and ST-PCA detect the location of anomalies accurately, although the proposed method was faster in detecting the hot-spot event (frame 53, whereas the ST-PCA method signaled at frame 63).

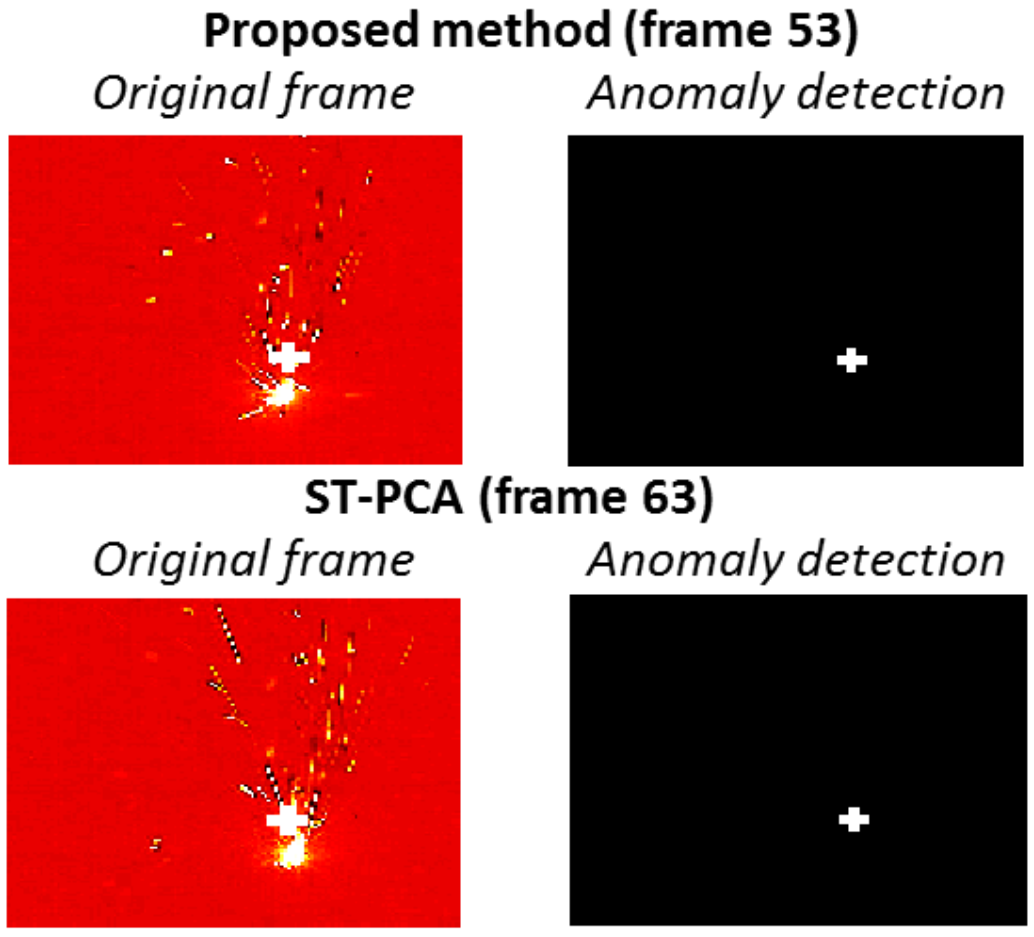

Figure 6 – Example of signals from different algorithms in one simulation runs; left: the original frames in correspondence of the first signaled alarm, right: the region of the frame signaled as the anomaly event, where the exact location is highlighted in white. (methods not shown in the figure did not signal in the considered simulation run).

### 3.3 Sensitivity Analysis

To understand how the tuning parameter selection $\lambda, \lambda_0$ and pre-processing procedure affects the result, Figure 8 shows different choices of $\lambda$ affect the run length of the medium size defect $(n = 20)$ in. We can clearly see that different choices of $\lambda$ does have an impact on the run length (RL) and localization accuracy (F-score). Too small values of $\lambda$ (e.g., $\lambda = 0.1$) lead to worse localization accuracy (F-score much smaller than 1) since too large weight is given to the most recent sample. However, too large values of $\lambda$ (e.g., $\lambda = 0.7$) lead to a good localization accuracy but also increases the detection delay since it does not give enough weight to the most recent samples. Similarly, too small values of $\lambda_0$ lead (e.g. $\lambda_0 = 0.1$) to a much larger variance of the algorithm due to the potential overfitting. Too large values of $\lambda_0$ (e.g. $\lambda_0 = 10{,}100$) greatly under-estimate the anomaly due to the shrinkage effect. In the present application, we advocate the implementation of the proposed approach to the original images with $\lambda = 0.3$ and $\lambda_0 = 1$. Calibration of the model parameters $\lambda, \lambda_0$ with the normal operation data is recommended since the different dynamic of foreground event may present.

Furthermore, we also study how the regularization parameters $\gamma_1, \gamma_2, \gamma_3$ would affect the results. We find that too small values of $\gamma_1$ (e.g. $\gamma_1 < 1e - 2$) could give too much flexibility of the foreground event estimation, and some anomalies and noise could be misspecified as the foreground event. However, too large values of $\gamma_1$ (e.g. $\gamma_1 > 1$) could also underestimate the foreground event, and these random foreground events could then lead to a worse estimation of the anomaly. For $\gamma_2$ , as long as it is set larger than 1, the detection delay and accuracy are not greatly influenced. Finally, the combined procedure of using multiple $\gamma_1$, $\gamma_2$ is suggested since it greatly reduces the RL and increases the F-value compared to using any $\gamma_1$ or $\gamma_2$ values. Finally, we find the algorithm is relatively robust to $\gamma_3$ unless $\gamma_3$ is too large. This is because too large values of $\gamma_3$ (e.g. $\gamma_3 > 0.01$) may put too much smoothness penalty on the anomaly, which leads to the under-estimation of small anomalies. Finally, we also investigated whether the LHZ removal pre-processing step can affect the performance of the proposed algorithm. Figure 7 (bottom panels) show that, when such a pre-processing step is applied, the proposed algorithm typically yields a larger RL. The reason is that the current implemented LHZ removal operation might sometimes accidentally remove the hot-spot too, which inflates the out-of-control detection delay. Despite this, the LHZ removal operation slightly increases the accuracy of the hot-spot localization, as a partial overlap between the LHZ and the hot-spot is avoided.

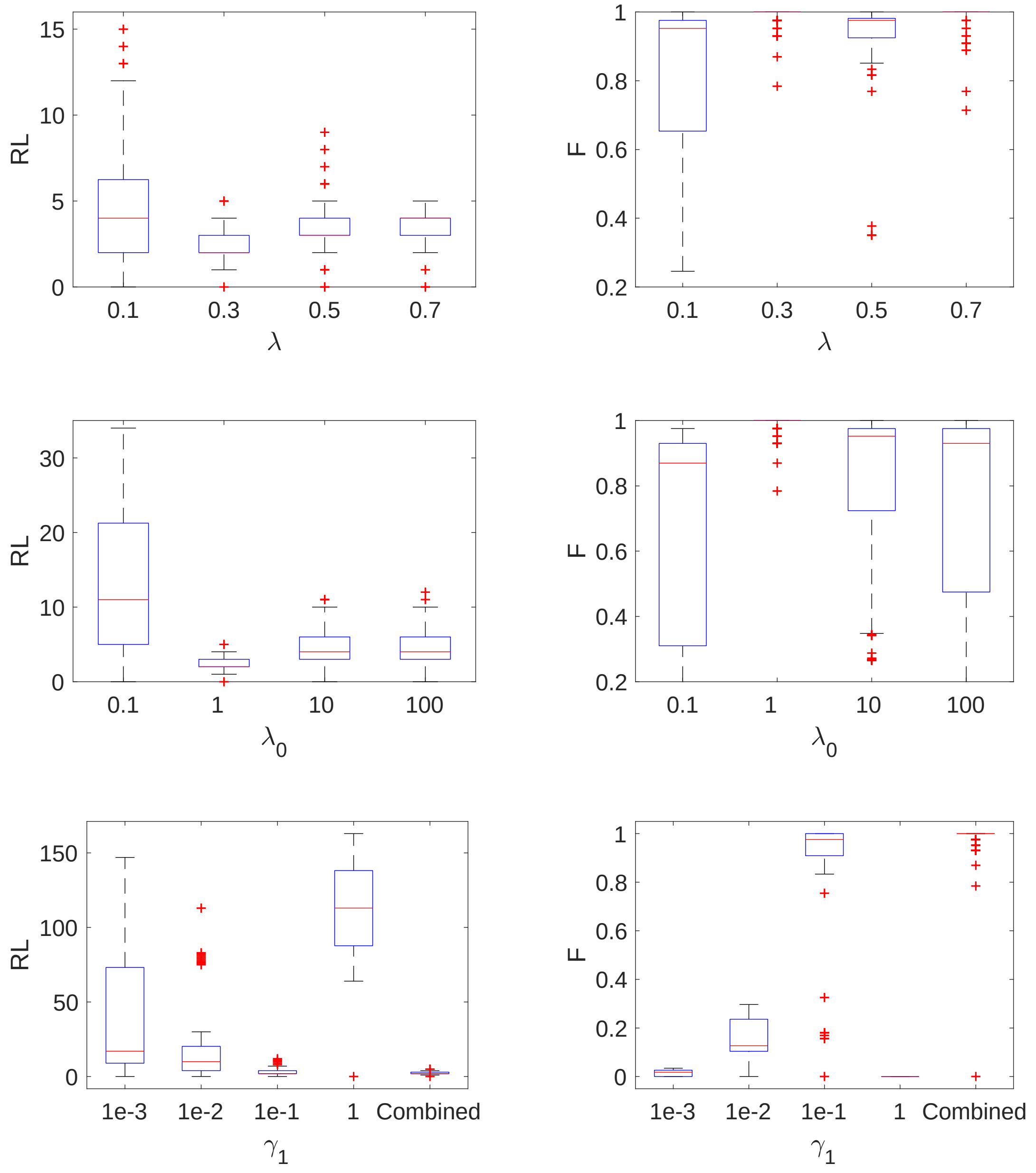

Figure 7 (Part A) - Sensitivity analysis of how $\lambda$ (1st row), $\lambda_0$ (2nd row), $\gamma_1$ (3rd row)

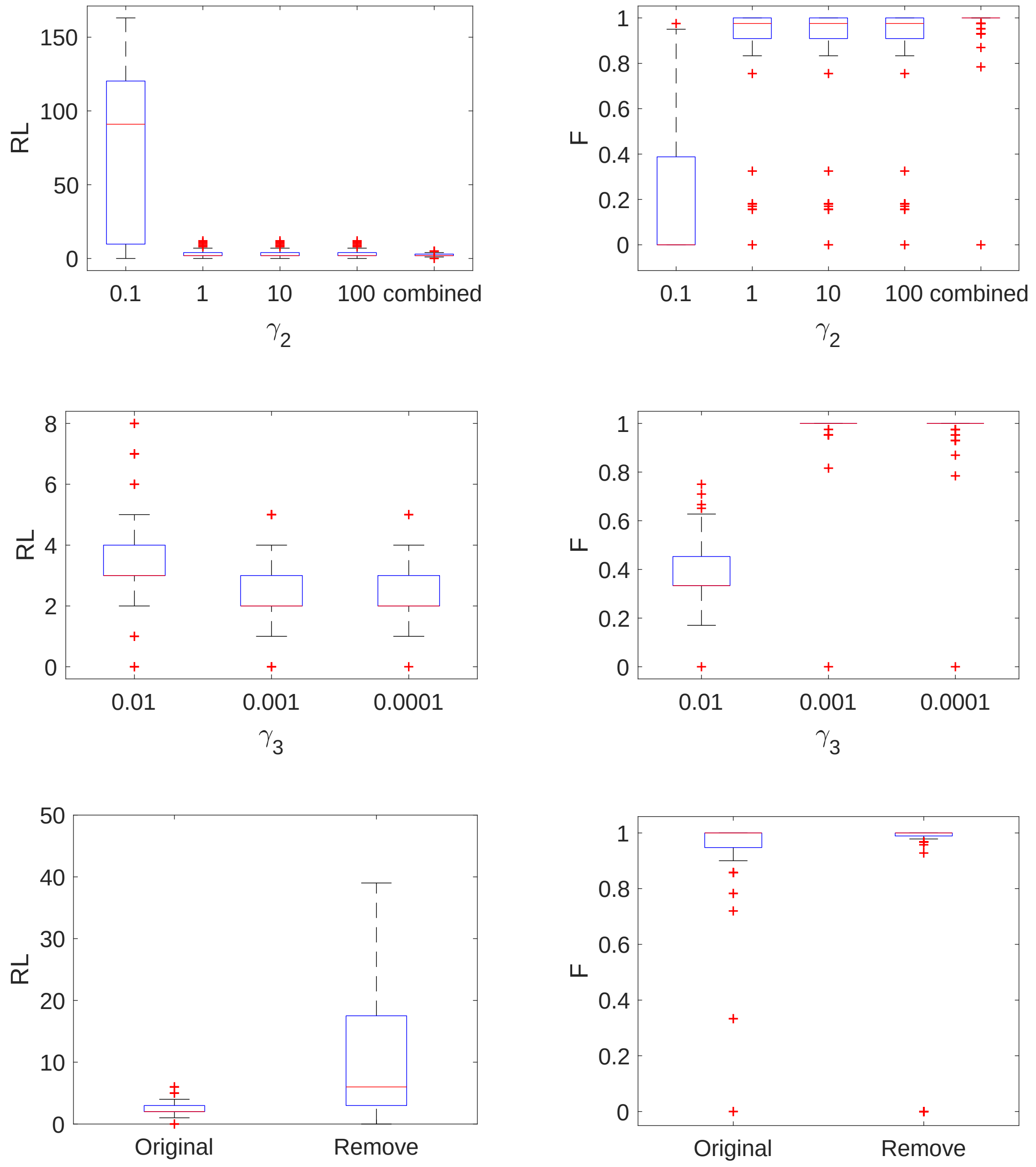

Figure 8 (Part B) - Sensitivity analysis of how $\gamma_2$ (1st row), $\gamma_3$ (2nd row), and the LHZ removal pre-processing step (3rd row) affect the RL and the F-score.

## 5. Case Study

In this section, we will evaluate the performance of the proposed algorithm using a case study during a real LPBF process with three different settings. More details of the experimental setup are discussed in Section 4.1. The proposed method is applied to the real case study with the comparison to other benchmark methods in Section 4.2.

### 4.1 Case Study Setup

The experimental case study previously presented in (Colosimo & Grasso, 2018) and (Grasso et al., 2017) was used to demonstrate the performances of the proposed approach in the presence of real hot-spot events. Previous studies showed that the occurrence of local over-heating conditions may yield geometrical distortions, especially in the presence of thin walls and acute corners. As a consequence of the local heat accumulation, surface tensions of the viscous melt cause the formation of solidified balls on the surface, leading to so-called super-elevated edges (Kleszczynski et al., 2012), i.e., ridges of the solidified material whose height may be higher than the layer thickness. Such local irregularities may propagate and inflate from one layer to another, with possible damages to the powder recoating system. This makes the quick detection of hot-spots particularly relevant in LPBF. As an example, Fig. 9 shows the consequence of hot-spots observed during the production of the complex shape used in this real case study in terms of geometrical irregularities in the part.

The same monitoring setup described in Section 4 was applied in this experimentation, but a higher sampling frequency was applied, i.e., $f = 300$ fps. This sampling frequency was selected as a compromise between the computational feasibility of in-situ image analysis and the ability to capture the process-related dynamic and transitory events without losing relevant information. The spatial resolution was about 150 µm/pixel. An image crop operation was applied to remove defocused regions and areas of the baseplate not involved by the process. The resulting image size was $121 \times 71$ pixels.

The experimentation consists of an LPBF process on AISI 316L powder with an average particle size of about $25 - 30\ \mu m$ for the production of the complex geometry shown in Figure 9 (overall dimensions of about $50\ \times 50\ \times 50\ mm$). In this case, three distinct videos were acquired during the LPBF of triangular-shaped slices in three different layers where hot-spots occurred in correspondence of acute corners in over-hang regions. This is caused by the fact that those regions are largely surrounded by loose powder, which has a lower heat transfer coefficient than the solid material. The hot-spot events produced local geometrical deformations in the printed part, as highlighted in Figure 9 (left panel). The LPBF specified with the process parameters summarized in Table 2.

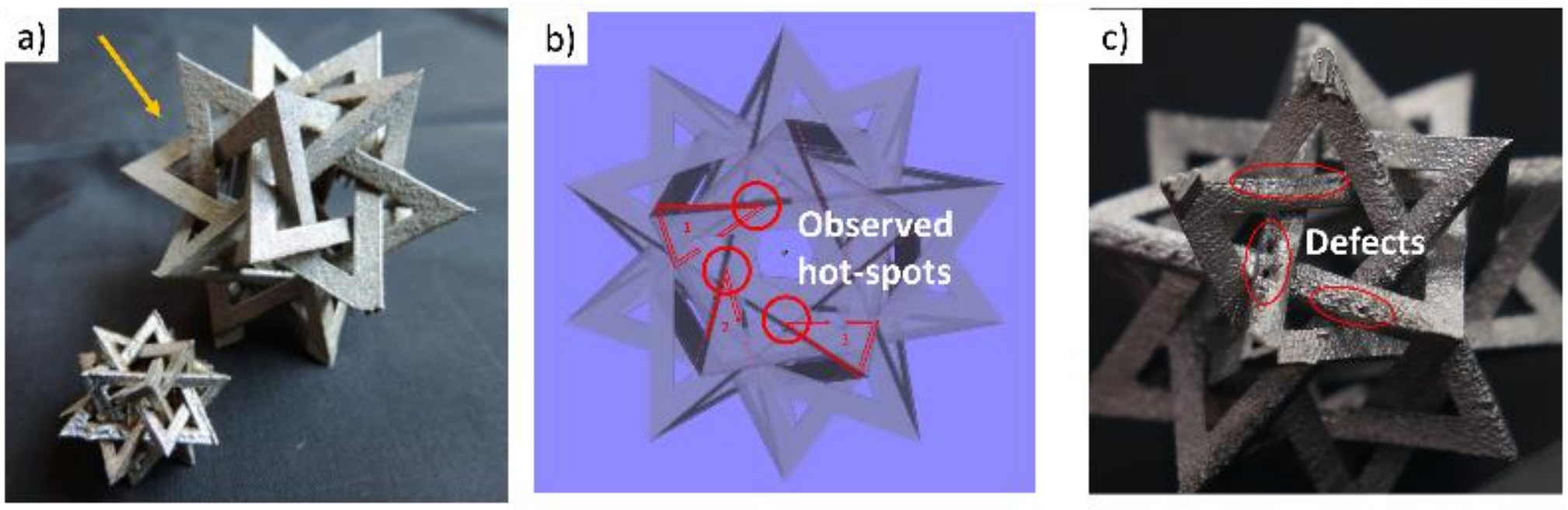

Figure 9. a) complex shape part used to test the proposed approach; b) examples of triangular portions of the sliced CAD model; c) local defects corresponding the acute corners of those triangles.

Table 2 – Main process parameters used in the experimental activity

| | Laser power ($P$) | Exposure time ($t$) | Focus position ($f_p$) | Point distance ($d_p$) | Hatch distance ($d_h$) | Layer thickness ($z$) |
|---|---|---|---|---|---|---|
| Value | $200\ W$ | $80\ \mu s$ | $0\ mm$ | $60\ \mu m$ | $110\ \mu m$ | $50\ \mu m$ |

The three video-sequences were labeled as Scenario A, B, and C, respectively. In each scenario, the laser beam passed over the hot-spot region more than once, and hence multiple events were sequentially observed. Different hot-spot events in the same video-sequence refer to the same location but different time intervals. Each time the laser scanned the defective area in correspondence of acute corners, the heat accumulation produced a hot-spot event that lasted for a few consecutive frames. Figure 10 shows some examples of videos frames acquired in the three scenarios. For each scenario, Figure 10 shows one example of video-frame under natural melting conditions (top-left panel) and video-frames corresponding to the begin and end of hot-spot events that were visible in the video sequence. At least four consecutive hot-spot events in the same location were visible in each scenario.

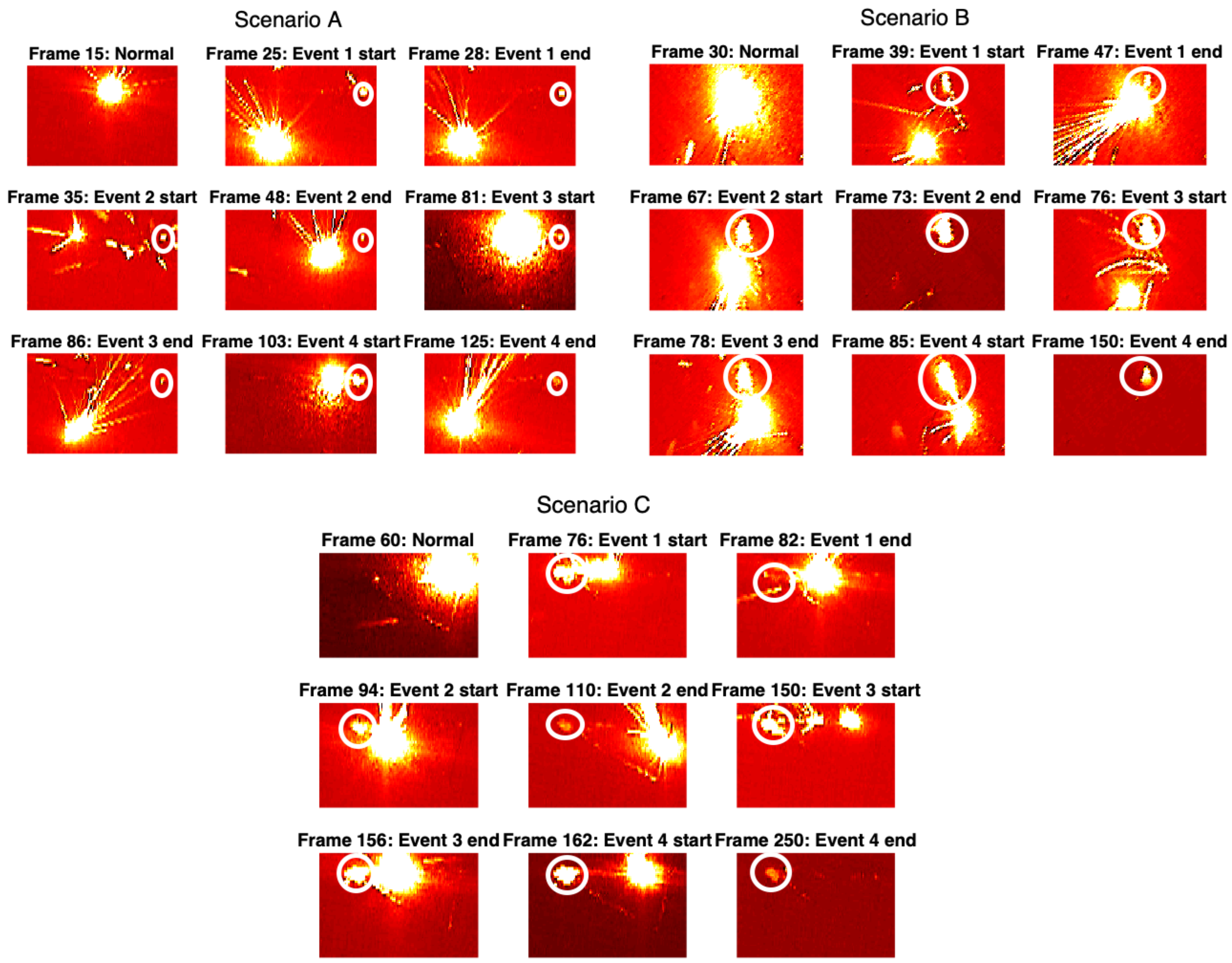

Figure 10. Example of frames from the three real-case scenarios where hotspot events are highlighted in white circle

Table 3. Performances indexes for different methods in different real case study scenarios

| Scenario | | | Time of first signal (frame index) | | | |
|---|---|---|---|---|---|---|
| | | | Event 1 | Event 2 | Event 3 | Event 4 |
| A | *Actual onset of hot-spot event* | | *25* | *35* | *81* | *103* |
| | Competing methods | Proposed | 25 | 38 | 84 | 107 |
| | | T-square | - | 40 | 84 | - |
| | | PCA | - | - | - | - |
| | | Lasso | - | - | - | 165 |
| | | STSSD | - | - | - | 167 |
| | | STPCA | - | 40 | 81 | 103 |
| B | *Actual onset of hot-spot event* | | *39* | *67* | *75* | *85* |
| | Competing methods | Proposed | 40 | 72 | 76 | 88 |
| | | T-square | - | - | 77 | - |
| | | PCA | - | - | 77 | - |
| | | Lasso | - | - | 77 | - |
| | | STSSD | - | - | - | - |
| | | STPCA | - | - | - | 94 |
| C | *Actual onset of hot-spot event* | | *77* | *94* | *150* | *162* |
| | Competing methods | Proposed | 78 | 97 | 152 | 164 |
| | | T-square | - | - | 152 | - |
| | | PCA | - | - | - | 168 |
| | | Lasso | - | - | 153 | - |
| | | STSSD | - | - | - | 168 |
| | | STPCA | - | - | - | 164 |

## 4.2 Performance Evaluation

Analogously to the simulation study, we applied all the competing methods to Scenarios A, B, and C, and the results are shown in Table 3. We evaluated the capability of competitor methods to detect the four major hot-spots events in each scenario by comparing the time of first signal (expressed in terms of frame index) with the time of first visible hot-spot occurrence in the image stream. When a method was not able to detect the hot-spot event, the symbol "-" was shown in Table 3. Table 3 shows that the proposed method properly detects all the hot-spot events. Following events were detected with no more than 5 frames of delay, which considerably outperforms all other competing techniques. $T^2$, PCA- Lasso- and STSSD-based methods were able to detect at most one real hot-spot event. The ST-PCA method was able to detect all the hot-spot events in Scenario A, apart from the very first one, with performances comparable to the ones provided by the proposed approach. However, in Scenario B and Scenario C, the ST-PCA approach allowed signaling only the last (and more severe) hot-spot events, with a larger delay than the proposed method. These results confirm that our proposed spatio-temporal methodology is more effective than the previously proposed ST-PCA technique, and it outperforms more traditional statistical methods for video-imaging analysis.

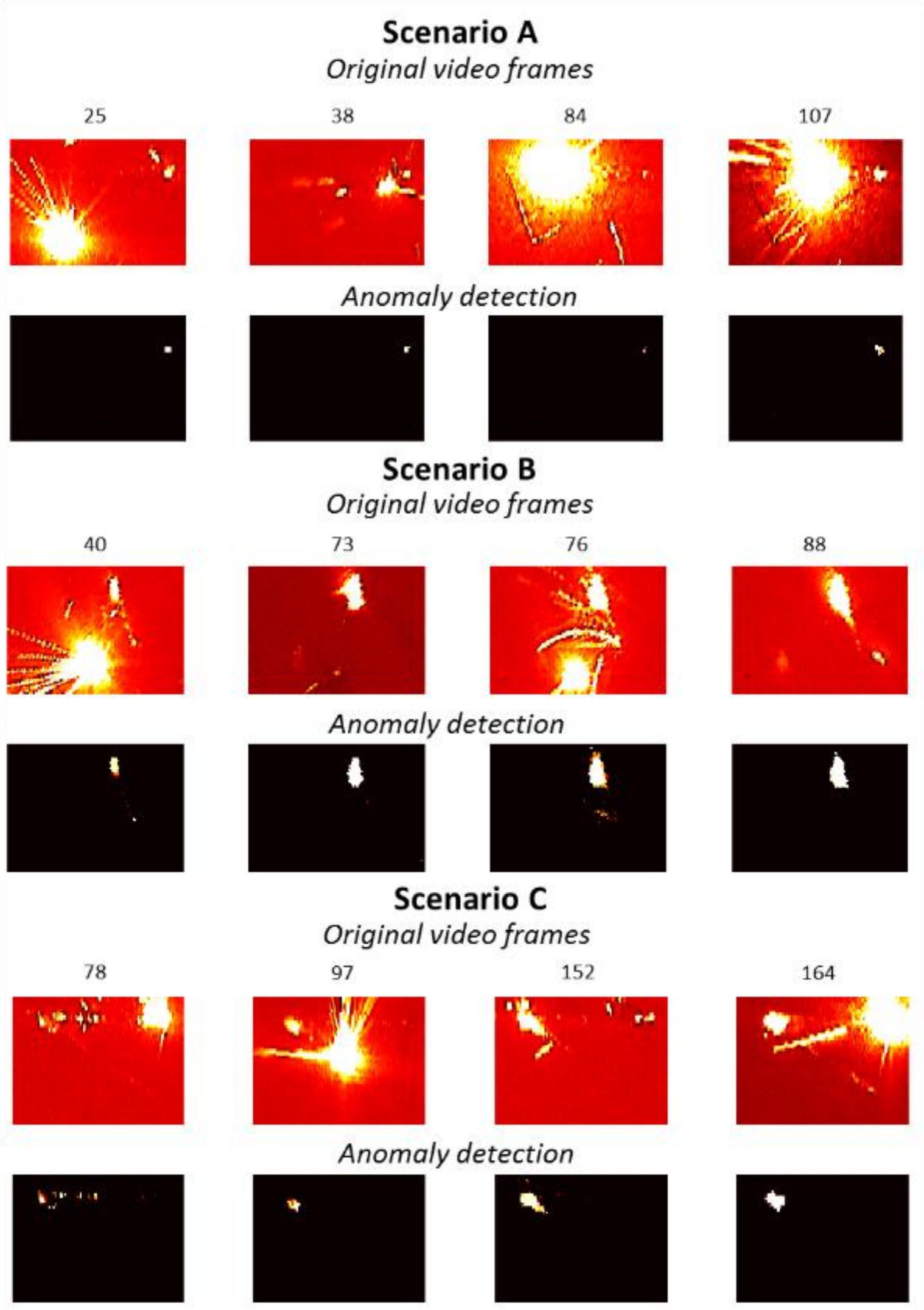

Figure 11. Detected anomalies for the proposed method in different scenarios; For each scenario, top 4 figures show the original images. Bottom 4 figures show the detected anomalies, where the anomaly location is shown in white.

Table 5 shows the computational time needed by each competing method (all of them were implemented in Matlab®) The proposed method requires 0.015s per sample with the approximation algorithm and 0.4s per sample without approximation, which is compliant with a real-time implementation. The $T^2$, PCA- and Lasso-based methods were even more computationally efficient than the proposed approach, but they provided much less accurate results. The ST-PCA method entails a recursive scheme that inflates the computational cost as new video frames become available. In all the real-case study analysis, the ST-PCA required less than 0.3 s at each iteration step. Thus, the proposed approach is not only more effective than the ST-PCA for hot-spot detection and localization but also more computationally efficient. In the present study, all methods ware tested in off-line mode, i.e., by first collecting real sensor data during the process and testing the proposed algorithm on these same data after the process. Nevertheless, the proposed method is designed for in-line and real-time use and satisfied the real-time implementation constraint.

Table 5. The computation time of all methods

| | Proposed | T-square | PCA | Lasso | ST-SSD | ST-PCA |
|---|---|---|---|---|---|---|
| Time | 0.01*s* | 0.0002 s | 0.004 s | 0.001 s | 0.006s | 0.28 s |
| Standard deviation | 0.0033s | 0.0004s | 0.001s | 0.0004s | 0.003s | 0.045s |

## 6. Conclusion

Online monitoring of high-dimensional spatio-temporal data is gaining an increasing interest in various manufacturing and other social applications. In this paper, we proposed a novel decomposition-based approach for real-time monitoring and anomaly detection of spatio-temporal data. Our method was specifically motivated by the "hot-spot" detection problem in metal additive manufacturing. The proposed method is able to take advantage of the layer-wise production paradigm to gather as much information as possible about the quality and stability of the process during the process itself, rather than (or in addition to) relying on traditional post-process quality controls. Most of these tools are mainly used to collect data during the process and provide the user with some post-process data reporting and/or datasets to support the investigation of specific problems and defects. What is missing in the industry is the availability of analytical tools able to quickly make sense of gathered data during the process and automatically signal the onset of defects and process instabilities.

The proposed method was tested in this framework to demonstrate the capability of overcoming the limitation of existing methods. Furthermore, the roposed method is general, and it can be applied to any image-based process monitoring applications where the foreground events are random and sparse, and the anomaly is spatially and temporally correlated. To handle the challenges of the high-dimensionality for the video-image stream, we proposed a recursive estimation procedure for real-time implementation of the algorithm. Finally, a sequential likelihood ratio test procedure was proposed for online change detection and anomaly localization. To demonstrate the effectiveness of the proposed method, we applied to both simulated and real data drawn from a real case study in LPBF. The results demonstrated that the proposed approach outperformed both methods previously presented for the same LPBF applications and other general techniques suitable to deal with video-imaging data. One promising direction for future research consists of extending the proposed method by incorporating a more complex time series modeling technique, e.g., the autoregressive-moving-average model (Hannan, 2009), to further enhance the characterization of the temporal structure embedded in spatio-temporal data. Moreover, future studies can be aimed at testing the proposed method in the presence of different kinds of out-of-control scenarios and different manufacturing applications.

## Appendix A – Soft-Thresholding

The weighted likelihood can be computed as

$$\begin{aligned}\operatorname{argmin}_{u_{s,t}} \sum_{t=1}^{T} \lambda^{T-t}\, l_t\left(\boldsymbol{\theta}, \{u_{s,t}\}_{s=1,\cdots,S}\right) \\ = \operatorname{argmin}_{u_{s,t}} \sum_{s} \left( \sum_{t=1}^{T} \lambda^{T-t} (\| x_{s,t} - \mu_{s,t} - u_{s,t} - \theta_s x_{s,t-1} \|^2 + \gamma ||u_{s,t}||_1) \right)\end{aligned}$$

Therefore, each $u_{s,t}$ needs to be optimized individually.

$$\hat{u}_{s,t} = \text{argmin}_{u_{s,t}} \parallel x_{s,t} - \mu_{s,t} - u_{s,t} - \theta_s x_{s,t-1} \parallel^2 + \gamma||u_{s,t}||_1$$

The optimality condition is given by:

$$0 \in \nabla \parallel x_{s,t} - \mu_{s,t} - u_{s,t} - \theta_s x_{s,t-1} \parallel^2 + \gamma\partial||u_{s,t}||_1$$

Then, $u_{s,t}$ can be solved by

$$u_{s,t} = S_{\frac{\gamma}{2}}(x_{s,t} - \mu_{s,t} - \theta_s x_{s,t-1})$$

## Appendix B - Derivation of the recursive estimation algorithm

The weighted likelihood can be computed as

$$argmin_{\boldsymbol{\theta}} \sum_{t=1}^{T} \lambda^{T-t}\, l_t\left(\boldsymbol{\theta}, \{u_{s,t}\}_{s=1,\cdots,S}\right)$$

$$= argmin_{\boldsymbol{\theta}} \sum_{s}\left(\sum_{t=1}^{T} \lambda^{T-t} \parallel x_{s,t} - \mu_{s,t} - u_{s,t} - \theta_s x_{s,t-1} \parallel^2 + \gamma_2 \parallel \theta \parallel_1 + \gamma_3 \parallel \theta \parallel_{TV} + \lambda_0 \parallel \theta \parallel^2\right)$$

$$= argmin_{\boldsymbol{\theta}} \sum_{s}\left(\left(\frac{1-\lambda}{1-\lambda^T}\sum_{t=1}^{T} \lambda^{T-t} x_{s,t-1}^2 + \lambda_0\right)\theta_s^2 - 2\theta_s \frac{1-\lambda}{1-\lambda^T}\left(\sum_{t=2}^{T} \lambda^{T-t} x_{s,t-1}\tilde{x}_{s,t}\right) + (\gamma_2 \parallel \theta \parallel_1 + \gamma_3 \parallel \theta \parallel_{TV})\right)$$

$$= argmin_{\boldsymbol{\theta}}\, \boldsymbol{\theta}^{\boldsymbol{T}}\, \widetilde{\Phi}_{S,T}^2 \boldsymbol{\theta} \quad - 2\boldsymbol{\theta}^T \widetilde{\boldsymbol{\Psi}}_T + \gamma_2 \parallel \theta \parallel_1 + \gamma_3 \parallel \theta \parallel_{TV}$$

$$= argmin_{\boldsymbol{\theta}} \parallel \widetilde{\boldsymbol{\Phi}}_{\boldsymbol{T}} \boldsymbol{\theta} - \widetilde{\boldsymbol{\theta}}_{\boldsymbol{T}} \parallel^2 + \gamma_2 \parallel \theta \parallel_1 + \gamma_3 \parallel \theta \parallel_{TV}$$

For simplicity, we define $\tilde{x}_{s,t} = x_{s,t} - \mu_{s,t} - u_{s,t}$, where $\widetilde{\boldsymbol{\theta}}_T = \left[\tilde{\theta}_{1,T}, \cdots, \tilde{\theta}_{S,T}\right]^T$, $\tilde{\theta}_{s,T} = \frac{\Psi_{s,T}}{\widetilde{\Phi}_{s,T}}$, $\widetilde{\boldsymbol{\Phi}}_{\boldsymbol{T}} = \text{diag}(\widetilde{\Phi}_{1,T}, \cdots, \widetilde{\Phi}_{S,T})$. $\widetilde{\boldsymbol{\Psi}}_T = \text{diag}(\widetilde{\Psi}_{1,T}, \cdots, \widetilde{\Psi}_{S,T})$.

$\widetilde{\Phi}_{s,T} = \sqrt{\frac{1-\lambda}{1-\lambda^T}\sum_{t=1}^{T}\lambda^{T-t}x_{s,t-1}^2 + \lambda_0}$ and $\widetilde{\Psi}_{s,t} = \frac{1-\lambda}{1-\lambda^T}\sum_{t=2}^{T}\lambda^{T-t}x_{s,t-1}\tilde{x}_{s,t}$. Here with the definition of $\Phi_{s,t} = \sum_{t=1}^{T}\lambda^{T-t}x_{s,t-1}^2$ and $\Psi_{s,t} = \sum_{t=2}^{T}\lambda^{T-t}x_{s,t-1}\tilde{x}_{s,t}$, we know $\widetilde{\Phi}_{s,T} = \sqrt{\frac{1-\lambda}{1-\lambda^T}\Phi_{s,t} + \lambda_0}, \widetilde{\Psi}_{s,t} = \frac{1-\lambda}{1-\lambda^T}\Psi_{s,t}$. Finally, we can easily verify that $\Phi_{s,t}$ and $\Psi_{s,t}$ can be updated recursively as

$$\Phi_{s,t} = \lambda\Phi_{s,t-1} + x_{s,t-1}^2, \Psi_{s,t} = \lambda\Psi_{s,t-1} + x_{s,t-1}(x_{s,t} - \mu_{s,t} - u_{s,t}).$$